\documentclass[runningheads]{llncs}

\usepackage{eccv}

\usepackage{eccvabbrv}

\usepackage{graphicx}
\usepackage{booktabs}
\usepackage{subcaption}
\usepackage[accsupp]{axessibility}  % Improves PDF readability for those with disabilities.

\usepackage{hyperref}

\usepackage{orcidlink}

\usepackage{algorithm}
\usepackage{algorithmicx}
\usepackage{algpseudocode}

\begin{document}

\title{Self-supervised Garment Dynamics with Persistent Wrinkles} 


\author{
Xiaoyuan Yang\inst{1}\orcidlink{0009-0001-1379-8918} \and
Deshan Gong\inst{3}\orcidlink{0009-0002-2516-9542} \and
Taku Komura\inst{3}\orcidlink{0000-0002-2729-5860} \and
He Wang\thanks{corresponding author, he\_wang@ucl.ac.uk}\inst{1,2}\orcidlink{0000-0002-2281-5679}
}

\authorrunning{X. Yang et al.}

\institute{
$^1$University of Leeds,
$^2$University College London,
$^3$The University of Hong Kong
}

\maketitle
\begin{center}
    \centering
    \captionsetup{type=figure}
    \includegraphics[width=\textwidth]{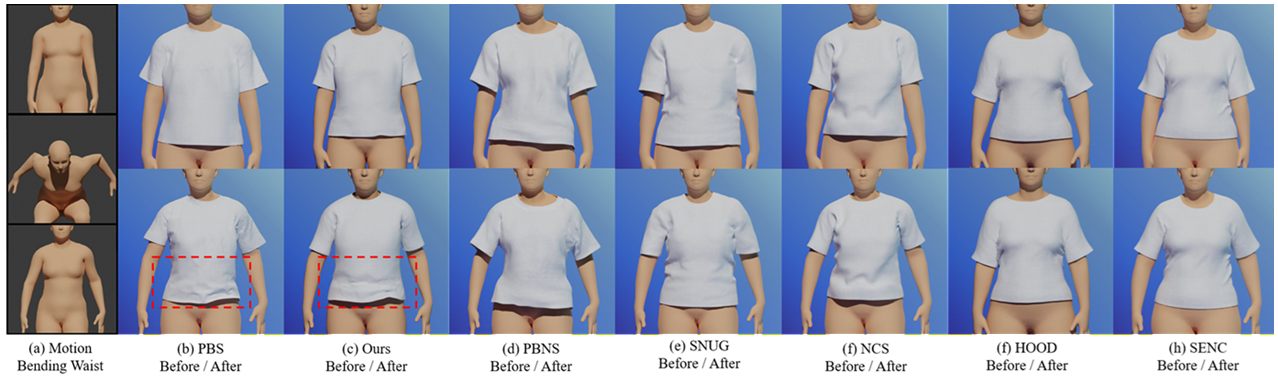}
    \captionof{figure}{Given the bending waist motion (a), the physics-based cloth simulator (PBS) can simulate persistent wrinkles around the abdomen area (b) caused by the material plasticity of fabrics. However, existing self-supervised cloth simulators (PBNS~\cite{bertiche2020pbns}, SNUG~\cite{santesteban2022snug}, NCS~\cite{bertiche2022neural}, HOOD~\cite{grigorev2023hood}, and SENC~\cite{liao2024senc}) naively model cloth as elastic material and cannot simulate persistent wrinkles (d, e, f, g, h). Our novel self-supervised simulator models material plasticity to simulate persistent wrinkles as the PBS (c).}
    \label{fig:teaser}
\end{center}%

\begin{abstract}
Self-supervised neural garment simulation has become popular due to its computational efficiency, good visual realism, and no reliance on training data. However, existing methods greatly simplify the mechanical properties of fabrics, ignoring persistent wrinkles caused by plasticity. Although this simplification allows for modeling of purely elastic material and simple training via energy minimization, the lack of believable wrinkles adversely affects the visual realism. Therefore, we introduce the first self-supervised neural garment simulator that explicitly models persistent wrinkles. This is accomplished through a novel physics-inspired loss function, which turns learning into a moving energy minimization problem to mimic plasticity. However, this requires learning to use a changing loss function, which causes difficulties in training because the loss function changes during optimization. To this end, we propose a new physics-inspired curriculum learning scheme where the target material for learning gradually changes from pure elasticity to elasto-plasticity, allowing the loss function and the learnable parameters to jointly converge. Through a comprehensive evaluation, we show that for the first time, self-supervised learning models can generate natural persistent wrinkles, outperforming existing methods on a variety of garments, body shapes, and body motions, according to a range of metrics. Our code is publicly available at \textit{https://github.com/realcrane/EPNet}
\end{abstract}

\section{Introduction}
Virtual garments are important digital assets for computer games, mixed reality, online marketing, \etc, where generating realistic virtual garments is universally required but labor-intensive. High-fidelity virtual garments are generated from physics-based simulation with the visual quality as the main goal~\cite{wang2021gpu}, but it is time-consuming and not suitable for interactive or real-time applications. In contrast, data-driven methods can quickly generate plausible garments, but they are not only less realistic but also strongly dependent on large amounts of training data~\cite{gundogdu2019garnet,gundogdu2020garnet++}. Recently, self-supervised neural garment simulators have emerged as a midway solution, where data reliance is eliminated and the time-consuming solution of physical systems is replaced by faster inference on neural networks.

Existing self-supervised learning (SSL) garment models focus on draping, cloth dynamics, collision handling, contacts~\cite{bertiche2020pbns,santesteban2022snug,bertiche2022neural}. Despite being designed for various purposes, these methods use a similar strategy for modeling bending. They often formulate a bending energy which is minimized during training. This strategy essentially treats the garment as an elastic material (at least in bending behaviors) and aims to train the network to predict meshes with a small bending energy. The formulation usually assumes a rest bending (RB) term, which is defined as the dihedral angle between edge-adjacent triangles~\cite{grinspun2003discrete}, and the deviation from this RB yields the bending potential energy. One example is $L_{bending} = \sum_{edges} k(\theta - \theta_{RB})^2$~\cite{bertiche2022neural} where $k$ is a scalar, $\theta$ is the predicted dihedral angle, and $\theta_{RB}$ is the dihedral angle in RB state. In existing methods, RB is often set manually. When setting RB to zero, it completely ignores plasticity, so the wrinkles generated by these methods are essentially left-over bending energy, and no sharp and persistent wrinkles can be generated; when setting RB to non-zero, it becomes \textit{dynamic}, which requires manual tuning for every edge in every frame. Furthermore, this dynamic change of RB should be physically based instead of being arbitrarily pre-defined, because plasticity on an edge can happen multiple times as the bending increases, manifested as \textit{multi-phasic plasticity}~\cite{gong2025cloth}. Consequently, the simplification of bending energy in existing methods leads to unrealistic garment deformation, failing to reproduce persistent wrinkles, a common phenomenon in garments. 

We propose the first SSL neural garment simulator that can generate natural persistent wrinkles. Given that wrinkles are mainly formed due to bending~\cite{narain2013folding,gong2025cloth}, we propose a new bending energy term based on dynamic RB. For every edge in the garment mesh, its RB is a spatio-temporal function of the deformation of its neighboring edges and its own deformation in the past. The design is inspired by that, spatially, existing methods can already predict the bending angles close to a pre-defined \textit{static} RB~\cite{bertiche2020pbns,santesteban2022snug,bertiche2022neural}. This gives us the confidence that neural networks can perform similarly well for dynamic RBs if they are known. However, since
dynamic RBs are not known \textit{a priori}, we unsupervisingly learn the RB as a time series with the correct dynamics, to mimic the multi-phasic plastic behavior of real fabrics. 

Despite the conceptual simplicity of our idea, training a neural network with dynamic RB presents an underexplored learning paradigm. Previously, the learning was a fixed energy minimization problem as the RB is pre-defined. Now it becomes a moving energy minimization problem as the RB is not known in the beginning of the training. This creates a chicken-egg problem: not knowing the RB makes it difficult to specify a loss function to train the model, but we can only compute the correct RB from the garment deformation predicted by a trained model. To this end, we propose a physics-inspired curriculum learning scheme which enables us to jointly learn the RB values and the model parameters.

We qualitatively and quantitatively evaluate our simulator and compare it with the state-of-the-art methods across various body motions, body shapes, and garment types, on a wide range of metrics. We demonstrate that, for the first time, persistent wrinkles can be generated from SSL models, which are both visually believable and physically plausible. Our contributions include:
\begin{enumerate}
    \item the first SSL neural garment model that can generate physically plausible and natural persistent wrinkles.
    \item a novel neural network, trained by energy minimization, containing a new dynamic bending energy term to mimic multi-phasic plasticity.
    \item a physics-inspired curriculum learning scheme to enable the training with changing loss functions.
\end{enumerate}

\section{Related Work}

\noindent\textbf{Physics-Based Cloth Simulation} 
Physics-based cloth simulation has been extensively studied since early deformable models~\cite{terzopoulos1987elastically}, with continuous advances toward visually realistic garment behavior~\cite{wang2011data, liang2019differentiable, li2022diffcloth, gong2022fine, gong2024bayesian, stuyck2022cloth}. Modern simulators incorporate sophisticated material models and fine-grained behaviors, including yarn-level effects and plastic deformation~\cite{kaldor2008simulating, cirio2014yarn, cirio2016yarn, narain2013folding, gong2025cloth}. 
While these methods can accurately model time-dependent and plastic phenomena through explicit physical integration, they are computationally expensive and unsuitable for real-time applications.

\noindent\textbf{Data-driven Garment Simulation} 
Data-driven approaches replace explicit simulation with neural networks trained on simulation data~\cite{wang2019learning, santesteban2019learning, gundogdu2019garnet, patel2020tailornet, tiwari2021deepdraper}. By learning deformation patterns from large datasets, these methods achieve fast runtime inference and can implicitly capture complex material behaviors present in the training data. However, they rely on extensive paired simulation datasets and their generalization ability is strongly tied to dataset coverage and diversity.

\noindent\textbf{Self-Supervised Garment Simulation} 
Self-supervised garment simulators remove the dependency on data by encoding physical priors into energy-based loss functions. PBNS~\cite{bertiche2020pbns} introduced the first self-supervised framework, formulating training as minimizing internal and external energy terms. Subsequent works improved architectural design and loss formulations: SNUG~\cite{santesteban2022snug} incorporated inertial effects, Neural Cloth Simulation~\cite{bertiche2022neural} decoupled static and dynamic feature learning, HOOD~\cite{grigorev2023hood} modeled local garment-body interactions with graph networks, ContourCraft~\cite{grigorev2024contourcraft} introduced contour-based interaction losses, SENC~\cite{liao2024senc} addressed self-collision through global intersection analysis, ESLR-Sim~\cite{liu2025extendedshortlongrangemesh} enhanced long-range information propagation, while Pb4U-GNet~\cite{liu2026pb4ugnetresolutionadaptivegarmentsimulation} addressed cross-resolution generalization. Despite the advances, these methods cannot simulate natural persistent wrinkles, which limits their visual realism. Our work introduces the first self-supervised neural garment simulator that generates natural persistent wrinkles.

\noindent\textbf{Wrinkle Simulation} 
Real-world cloth develops wrinkles when deformed, and realistically simulating wrinkles is essential for visual plausibility~\cite{benusiglio2012anatomy}. Realistic wrinkle simulation has long been a goal in computer graphics. Cloth wrinkles are commonly categorized as dynamic or static~\cite{larboulette2004real}. Dynamic wrinkles refer to transient folds that form and disappear during motion. In contrast, static wrinkles are permanent deformations that persist even after the cloth is no longer under load. Early efforts in simulating dynamic wrinkles employed geometric models to define cloth fold propagation~\cite{aono1990wrinkle}. To achieve greater physical realism, later research integrated physics into wrinkle formation modeling~\cite{bridson2003simulation}. 

More recently, data-driven approaches have learned wrinkles from data~\cite{wang2010example} or use generative models to add folds to garments~\cite{lahner2018deepwrinkles}. In contrast, research on simulating static wrinkles has been underexplored. To model permanent wrinkles, \cite{narain2013folding} employed a hardening plastic model. \cite{miguel2013modeling} used the Dahl friction model to capture hysteresis and simulate unrecoverable wrinkles. More recently, \cite{gong2025cloth} introduced a time-dependent friction and plastic model to simulate time-dependent wrinkles. Our work focuses on static wrinkle simulation. Unlike prior work, our model is a self-supervised neural simulator that enables fast inference without requiring training data.

\section{Method}

Our simulator is designed for simulating on-body garments. It models a garment as a thin-shell which is discretized into a triangle mesh of $N$ vertices and $D$ edges. Its state $\mathcal{S}$ consists of vertex positions $\mathbf{x} \in \mathbb{R}^{3 N}$ and velocities $\dot{\mathbf{x}} \in \mathbb{R}^{3 N}$. Garment motions are denoted by $\{\mathcal{S}^{(t)} | t \in \mathbb{Z}, t \in [0, T]\}$, where $\mathcal{S}^{(t)}$ is the garment state at time step $t$. For the body skeleton with $j$ joints, we denote its state by $\mathcal{B} \in \mathbb{R}^{9 j}$ where each joint has a feature vector in $\mathbb{R}^9$ denoting its location and orientation~\cite{bertiche2022neural}. To represent the human body surface with respect the skeleton, we use SMPL~\cite{loper2015smpl}.

To simulate a garment, our simulator predicts the future garment states, given its initial state and a body motion: $\mathcal{S}^{(1:T)} = f_{sim}(\mathcal{S}^{(0)}, \mathcal{B}^{(0:T)}; \boldsymbol{\Theta})$ where $f_{sim}$ denotes our self-supervised garment simulator with learnable parameter $\boldsymbol{\Theta}$. $\boldsymbol{\Theta}$ is learned by minimizing a physics-based loss function:
\begin{equation}
    \mathcal{L} = \sum_{t}^T \frac{h^2}{2} (\frac{\frac{\mathbf{x}^{(t+1)} - {\mathbf{x}}^{(t)}}{h} - \dot{\mathbf{x}}^{(t)}}{h})^\top \mathbf{M} (\frac{\frac{{\mathbf{x}}^{(t+1)} - {\mathbf{x}}^{(t)}}{h} - \dot{\mathbf{x}}^{(t)}}{h}) + W^{(t)}
    \label{eq:loss_all}
\end{equation}
where $\mathbf{M}$ is the garment lumped mass matrix, $W$ denotes the potential energy due to, \eg, gravity, garment stretching energy, bending energy, \etc, and $h$ is the time step size. This minimization is equivalent to using implicit Euler~\cite{baraff2023large} to solve the equation defined by the Newton's Second law: $\mathbf{M}\ddot{\mathbf{x}}=\mathbf{f}=-\nabla W$, where $\mathbf{f}$ is the resultant force imposed on the garment~\cite{martin2011example}. The potential energy $W$ contains several terms: 
\begin{equation}
    W = w_{b} W_{bend} + w_{st} W_{stretch} +w_{sh} W_{shear} + w_{c} W_{collision} + W_{gravity}
    \label{eq:energies}
\end{equation}
where $W_{bend}$, $W_{stretch}$, $W_{shear}$, $W_{collision}$, and 
$W_{gravity}$ denote the bending, stretching, shear, collision, and gravitational energy terms. Among them, $W_{bend}$ determines the bending mechanics which dictates the persistent wrinkle formation. Therefore $W_{bend}$ is the key. 

\subsection{Physics-inspired Bending Energy for Wrinkles}

In physics-based garment simulations, a common way to simulate persistent wrinkles is modeling garments as elasto-plastic material. In this case, the total bending is decomposed into elastic and plastic deformation: $\varepsilon = \varepsilon_e + \varepsilon_p $, where $\varepsilon$, $\varepsilon_e$ and $\varepsilon_p$ denote the total bending strain, elastic bending strain, and plastic bending strain, respectively. The elastic bending stress always tends to keep the garment in the state where $\varepsilon_e = 0$. Therefore, when $\varepsilon_p=0$, the elastic bending stress keeps the garment in the shape with zero strain.  $\varepsilon_p$ is only updated when $\varepsilon_e$ becomes greater than the yield strain, \ie, $| \varepsilon_e | > \varepsilon_y$ where $\varepsilon_y$ is the yield strain serving as a threshold. 

$\varepsilon_p$ is the RB in existing methods which is predefined and static. Unlike existing methods, our RB is self-supervisedly learned and dynamic, by adopting the perfect plastic model~\cite{o2002graphical} where the part of the elastic strain exceeding the yield strain is immediately converted to the plastic strain:
\begin{equation}
    \varepsilon_p \rightarrow \varepsilon_p + \text{sign}(\varepsilon_e) (| \varepsilon_e | - \varepsilon_y)
    \label{eq:plasticStrain}
\end{equation}
Note $\varepsilon_p$ can change over time and is a function of the past plastic strains for an edge. In practice, we compute \cref{eq:plasticStrain} by:
\begin{align}
    \Delta\varepsilon^{(t)} &= \varepsilon^{(t)}-\varepsilon^{(t-1)}_{p}-\varepsilon_y \label{eq:delta_plastic} \\
    \varepsilon^{(t)}_p &= \varepsilon^{(t-1)}_p+\text{sigmoid}(k_p\Delta\varepsilon^{(t)})\Delta\varepsilon^{(t)} \label{eq:epsilon_plastic}
\end{align}
where $\varepsilon^{(t)}$ is the total strain of an edge at time $t$. $\varepsilon^{(t)}_{p}$ is the plastic strain at time $t$. $k_p$ is a coefficient. Finally, our bending energy is computed as:
\begin{equation}
    W_{bend}  = \frac{1}{TD} \sum_{t}^{T} \sum_{d}^{D} k_b \frac{l^2}{8a} (\varepsilon^{(t)}_d - \varepsilon^{(t)}_{p, d})^2 \label{eq:bending_loss}
\end{equation}
where $k_b$ is the bending stiffness, $l$ is the length of the edge, $a$ is the sums of the areas of the two triangles, $d$ is the edge index.
$\varepsilon^{(t)}_d$ and $\varepsilon^{(t)}_{p, d}$ are the total strain and the RB at time $t$ for edge $d$. $\varepsilon^{(t)}_{p, d}$ is defined as the bending angle at which the bending energy is zero, corresponding to the dihedral angle between two adjacent triangles. 

The introduction of the new $W_{bend}$ creates a chicken-egg problem due to that the RB $\varepsilon^{(t)}_{p, d}$ is unknown. In previous methods, $\varepsilon^{(t)}_{d}$ is predicted by the model and $\varepsilon^{(t)}_{p, d}$ is predefined, so the learning is a fixed energy minimization problem. However, in our method, $\varepsilon^{(t)}_{p, d}$ needs to be computed based on \cref{eq:epsilon_plastic,eq:delta_plastic} which needs $\varepsilon^{(t)}_{d}$, but the model needs to be trained well to predict $\varepsilon^{(t)}_{d}$. The inter-dependence between $\varepsilon^{(t)}_{d}$ and $\varepsilon^{(t)}_{p, d}$ requires us to solve a moving energy minimization problem. 

\subsection{Physics-inspired Curriculum Learning}

To solve this chicken-egg problem, we introduce two networks: an Elastic Network (E-Net) and a Plastic Network (P-Net). E-Net simulates garment elastic dynamics, and P-Net predicts the evolving RB to model the plastic deformations formed in motions. 

\begin{equation}
    \mathcal{S}^{(1:T)} = \mbox{E-Net} (\mathcal{S}^{(0)}, \mathcal{B}^{(0:T)}, \varepsilon_{p, d}^{(0:T)}); \quad
    \varepsilon_{p, d}^{(t)} = \mbox{P-Net} (\mathcal{S}^{(t-1)}, \varepsilon_{p, d}^{(t-1)})
    \label{eq:nets}
\end{equation}
E-Net and P-Net are mutually dependent: the $\varepsilon_{p, d}^{(0:T)}$ of the E-Net in \cref{eq:nets} dependents on output of P-Net, and the input of the P-Net, $\varepsilon_{p, d}^{(t-1)}$, depends on the output of E-Net. Theoretically, to train our simulator, we could initialize both E-Net and P-Net and rely on optimization to lead to a convergence. In practice, the convergence proves to be difficult. So we break the tie, and train E-Net and P-Net alternatively. Furthermore, since we need to enforce \cref{eq:delta_plastic,eq:epsilon_plastic} during training, \ie, the P-Net prediction needs to satisfy both equations between frames, the design of the alternative training scheme needs special care. To this end, inspired by curriculum learning~\cite{wang2021survey}, we design a curriculum where the physical behaviors of the garment change from simple to complex, which corresponds to changing from purely elastic to elasto-plastic material.

We denote the current iteration between E-Net and P-Net by another superscript $i$ and represent the RB of an edge in time $t$ as $\varepsilon^{(t, i)}_p$. We keep both $t$ and $i$ to distinguish between the frame and the training iteration. The whole curriculum learning is shown in \cref{alg:CL}. At the high level, we first use a near-zero RB to train E-Net as if the garment is purely elastic. Then we update the RB from the E-Net prediction via \cref{eq:delta_plastic,eq:epsilon_plastic}. Next, the updated RB is used to train P-Net as the target RB. The training is done by minimizing the mean squared error between the predicted RB by P-Net and the target RB. Then the whole process repeats itself until convergence. As the RB accumulates during training when large deformation is present, the material gradually changes from purely elastic to elasto-plastic.

\label{sec:cl}
\begin{algorithm}[bt]
\caption{Curriculum Learning of Self-supervised Elasto-Plastic Material}
\label{alg:CL}
\begin{algorithmic}[1]
\State Initialize $\varepsilon^{(t, 0)}_p, \varepsilon_p \sim \mathbf{N(0, \sigma)}$ , $\varepsilon_y$ = \textit{user defined}
\For{$i = 1, 2, \dots, I$}
    \State Train E-Net via \cref{eq:loss_all} with $\varepsilon_p$ as the RB
    \State Predict $\varepsilon^{(t)}$ for every edge using E-Net
    \For{$t = 1, 2, \dots, T$}
        \State Compute $\varepsilon^{(t, i)}_p$ for every edge via \cref{eq:delta_plastic,eq:epsilon_plastic}. 
    \EndFor
    \State $\varepsilon_p \leftarrow \varepsilon^{(t, i)}_p$
    \State Train P-Net with $\varepsilon_p$ as the target RB
    \State Predict $\varepsilon'_p$ using P-Net
    \State $\varepsilon_p \leftarrow \varepsilon'_p$
\EndFor
\State \Return trained model parameters $\boldsymbol{\Theta}$
\end{algorithmic}
\end{algorithm}

\subsection{Neural Elastic and Plastic Model}

Our novel curriculum learning method is highly flexible and imposes no architectural restrictions on the design of E-Net and P-Net. While the specific instantiations of E-Net and P-Net used in this work are inspired by prior self-supervised simulators for their simplicity and ability to capture garment mechanics, our learning method is theoretically architecture-agnostic. It can be used in a plug-and-play manner to simulate elasto-plastic garments that exhibit persistent wrinkles.

\begin{figure*}[tb]
    \centering
    \includegraphics[width=\linewidth]{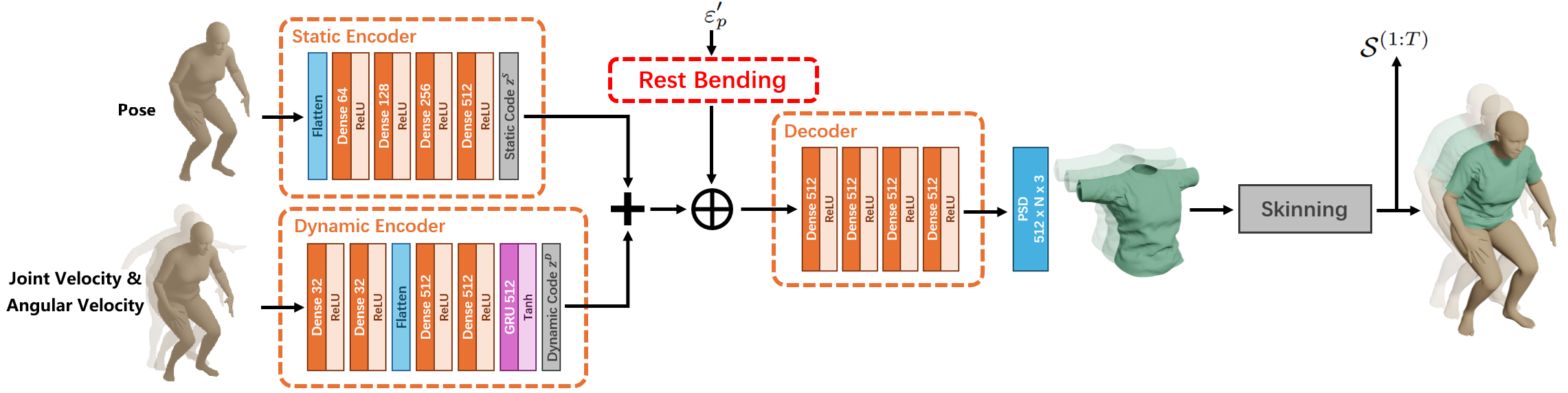}
    \caption{Elastic Net (E-Net): The input consists of a pose and its first-order time derivative. After passing through the encoder, the RB predicted by P-Net is used as a conditional input to the decoder, which then predicts the final garment deformation $\mathcal{S}^{(1:T)}$. Based on the predicted deformation, the RB is computed and subsequently used for P-Net training.
}
    \label{fig:elastic_net}
    \vspace{-1em}
\end{figure*}

\noindent\textbf{Elastic Net} 
Similar to~\cite{bertiche2022neural}, we adopt a recurrent encoder-decoder to model body motion in two subspaces: static subspace and dynamic subspace, shown in~\cref{fig:elastic_net}. We use the Static Encoder to encode the body static pose and the Dynamic Encoder to handle body motions. The outputs of the two encoders are combined and concatenated with the RB, which is predicted by P-Net (on the first training iteration, RB is initialized to a near-zero value). The combined features are then fed into the Decoder to predict garment deformations for all frames. The predicted deformations are applied through skinning to fit the garment onto the body. Based on the simulated deformation, RB is updated by using \cref{eq:delta_plastic,eq:epsilon_plastic}, which is used as the training target RB for P-Net.
 
\noindent\textbf{Plastic Net} 
P-Net adopts an encode--process--decode message-passing architecture~\cite{grigorev2023hood} defined on an intrinsic hinge graph. Its goal is to predict the evolving RB, which captures the accumulated plastic deformation in the garment. 
Persistent wrinkles exhibit both spatial and temporal correlations: spatially, wrinkles propagate across neighboring mesh regions, while temporally plastic deformation accumulates over successive bending motions. We model these correlations using a hinge-graph message-passing network with incremental updates over time.

\begin{figure*}[tb]
    \centering
    \includegraphics[width=\linewidth]{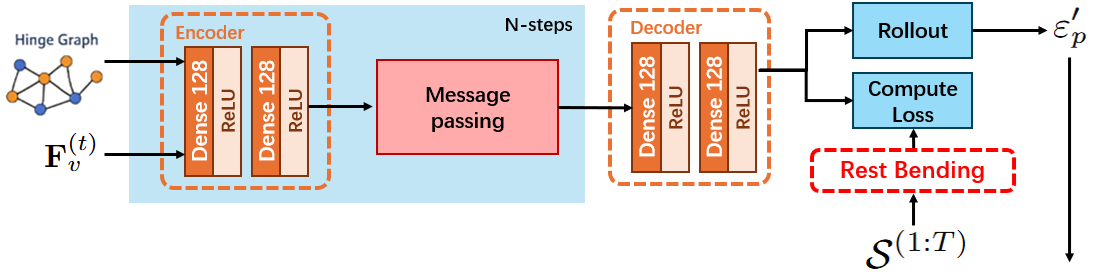}
    \caption{Plastic Net (P-Net): Hinge-level node and edge features at time step t are first embedded by MLP encoders. We then apply 3 message-passing steps, where edge features are updated using an edge MLP 
    $f_{v\to e}(e_{ij}, v_i, v_j)$ and node features are updated using a node MLP 
    $f_{e\to v}(v_i, \mathrm{mean}_j(e_{ij}))$. After decoding, the Rest Bending (RB) computed by \cref{eq:delta_plastic,eq:epsilon_plastic} from the E-Net predicted deformation $\mathcal{S}^{(1:T)}$ as the training target for learning. By rolling out P-Net over time produces the predicted  $\varepsilon'_p$ which is used as RB for condition E-Net.}
    \label{fig:plastic_net}
    \vspace{-1em}
\end{figure*}

To capture spatial correlations of plastic deformation, we construct a hinge graph $\mathcal{G}=(\mathcal{V},\mathcal{E})$ derived from the garment mesh. Each node $v\in\mathcal{V}$ corresponds to an interior mesh hinge (a pair of adjacent triangles sharing an edge), and edges $(v_i,v_j)\in\mathcal{E}$ represent adjacency between neighboring hinges. Bending angles are defined per hinge, and message passing is performed on this intrinsic hinge graph. As shown in \cref{fig:plastic_net}, at each time step P-Net takes hinge-level node and edge features $\mathbf{F}_v^{(t)}$ derived from local geometry (deformed and rest configurations), timestep information, and material parameters. These features are embedded by MLP encoders and processed using three steps of message passing. The MLP decoder outputs a scalar increment for each hinge, representing the predicted update of the RB.

Persistent wrinkles form when the bending along an edge exceeds a certain threshold, establishing a non-zero RB. Once formed, the fabric tends to return to this RB configuration rather than the flat state. If further deformation occurs and exceeds another threshold, RB increases and the wrinkle becomes sharper. Consequently, the RB of each hinge depends on its previously accumulated state.
To model this temporal evolution, P-Net predicts incremental plastic updates at each time step (via \cref{eq:delta_plastic,eq:epsilon_plastic}). Rolling out these updates over time produces the predicted RB sequence, which is then used to condition E-Net during the next training iteration. Unlike existing self-supervised garment simulators, which primarily model transient garment dynamics through recurrent state propagation, our method explicitly maintains and updates a persistent rest-bending state, enabling the accumulation of irreversible wrinkle effects over long-term deformation histories.

\section{Experiments}

\noindent\textbf{Data} We use the widely employed AMASS dataset~\cite{AMASS_ACCAD}, which includes the SMPL body model, the MANO hand model (52 joints, 6,890 vertices), and 252 motion sequences with approximately $188,928$ poses. Following the same setting of existing methods~\cite{bertiche2022neural}, we use $90 \%$ data for training, $10 \%$ for testing. Before training, we extract the joints for representing body poses~\cite{bertiche2022neural}. 

\noindent\textbf{Baselines} We employ a PBS with a perfect plastic model~\cite{narain2013folding} as the ground-truth. To highlight our model's unique ability to simulate persistent wrinkles, we also compare our model with the state-of-the-art self-supervised neural cloth simulators: PBNS~\cite{bertiche2020pbns}, SNUG~\cite{santesteban2022snug}, NCS~\cite{bertiche2022neural}, HOOD~\cite{grigorev2023hood}, and SENC~\cite{grigorev2024contourcraft}. 

\noindent\textbf{Evaluation Motions} Since there is no ground-truth in self-supervised learning, we employ PBS with plasticity as reference for evaluation. However, we found that not all motions in AMASS are suitable for evaluation. To ensure meaningful wrinkle assessment, we exclude motions that produce negligible deformation which does not lead to persistent wrinkles. We also exclude motions that produce unresolvable penetrations in PBS. At the end, we choose six representative motions (bending waist, ascending stairs, dancing, dodging, jumping, raising hand) for physical simulation to generate reference garments. 

\noindent\textbf{Metrics} We adopt Bending Error Distribution (BED), Plastic Error (PE), Bending Error (BE), Mean Euclidean Distance (MED), Chamfer Distance (CD), Bending Energy (BEN) for comparison (all lower is better), and garment-body collision (COL). 
BED is computed on the top 10\% edges selected by Segmented Maximum Analysis based on peak plastic bending over the sequence, focusing the evaluation on wrinkle-relevant regions. PE is the per-edge $\mathrm{L1}$ error of the  $\varepsilon_p'$ predicted by P-Net. Note that PBNS, SNUG, NCS, HOOD, and SENC do not have RB so their RB is effectively zero. BE is per-edge $\mathrm{L1}$ error of the predicted $\varepsilon$ from E-Net. MED and CD is the mean Euclidean distance and Chamfer Distance between the predicted vertex locations and the ground-truth. BEN is the mean per-edge bending energy. COL is the mean garment-body penetration depth.
Full metric definitions are provided in the supplementary material.

\subsection{Persistent Wrinkles on Unseen Motions}

\begin{table}[tb]
    \centering
    \caption{
    Numerical comparison on unseen motions. Metrics are averaged across all testing motions.
    }
    \resizebox{\textwidth}{!}{
    \begin{tabular}{lccccccc}
        \toprule
Base/Mtr & BED(rad)$\downarrow$ & PE(rad)$\downarrow$ & BE(rad)$\downarrow$ & MED($m$)$\downarrow$ & CD($m^2$)$\downarrow$ & BEN$\downarrow$ & COL($m$)$\downarrow$  \\
        \midrule
PBNS & 0.60223 & 0.12506 & 0.18733 & 0.04038 & 0.00054 & 0.04565 & \textbf{0.00095} \\
SNUG & 0.64567 & 0.12506 & 0.20327 & 0.03191 & 0.00061 & 0.05976 & 0.00136 \\
NCS  & 0.76556 & 0.12506 & 0.17611 & 0.03210 & 0.00040 & 0.03412 & 0.00129 \\
HOOD & 0.53553 & 0.12506 & 0.20948 & 0.06081 & 0.00118 & 0.13150 & 0.00210 \\
SENC & 0.52529 & 0.12506 & 0.19785 & 0.05486 & 0.00074 & 0.13657 & 0.00160 \\
Ours & \textbf{0.28714} & \textbf{0.03206} & \textbf{0.17118} & \textbf{0.02355} & \textbf{0.00027} & \textbf{0.02999} & 0.00129 \\
        \bottomrule
    \end{tabular}
    }
    \label{tab:gen_motions}
\end{table}

We show a visual comparison in \cref{fig:teaser}. Despite all method generate wrinkles, our method is the most similar to the PBS, especially on the front of the t-shirt. PBNS, HOOD, and SENC generates few wrinkles. SNUG and NCS generate sharp wrinkles but in the wrong places. Most of the sharp persistent wrinkles should be around the waist as the motion is waist bending as shown in \cref{fig:teaser} (a). Furthermore, our method captures the causality clearly. Before bending, wrinkles should not appear around the abdomen area while sharp persistent wrinkles should appear after the bending. However, PBNS, SNUG, NCS, HOOD, and SENC do not show this causality and generate wrinkles even before bending.

We further show numerical comparison in \cref{tab:gen_motions}. The smaller BED, PE, BE, and BEN directly demonstrate that our method is closer to PBS when forming persistent wrinkles. The learned plasticity reduces the potential energy stored in the deformation hence lower energy and error. This also leads to more accurate geometries indicated by smaller MED and CD. Our method is universally better across metrics. For each metric, comparing with the best-performing baseline, our approach improves the performance by 45.34\% in BED, 74.36\% in PE, 2.79\% in BE, 26.19\% in MED, 32.50\% in CD, and 12.10\% in BEN, while achieving the second-best COL, indicating that the improved wrinkle modeling does not compromise collision handling.

\subsection{Wrinkles on Unseen Bodies and Motions}

\begin{figure*}[tb]
    \centering
    \includegraphics[width=\textwidth]{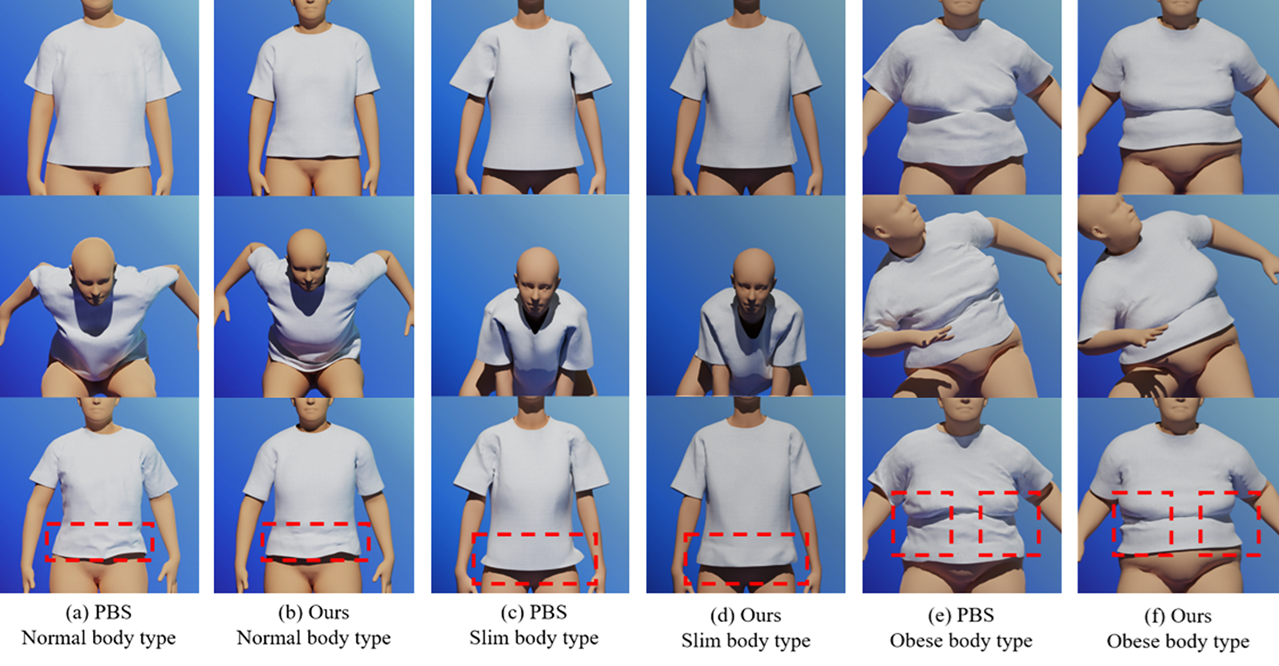}
    \caption{Visual comparison between our simulator and PBS across unseen bodies and motions: slim, and obese. Trained on the normal body, our simulator can generate wrinkles that are closely match PBS (a,b). Furthermore, it can generate plausible wrinkles on the slim and obese body which are unseen in training. The positions of the formed wrinkles are similar to PBS as well (c,d,e,f).}
    \label{fig:diff_body}
\end{figure*}

Once trained, our P-Net can generalize to different bodies, as it can predict the RB based on motions and garment itself. \cref{fig:diff_body} compares the wrinkles simulated by our model and PBS when the t-shirt is on three different bodies: Normal, Slim, and Obese, where the latter two are not used in training. Also, all the motions are unseen at the training stage either. Trained well on the Normal body \cref{fig:diff_body} (b), the wrinkles on the Slim and Obese body (\cref{fig:diff_body} (c-f)) also look similar to the PBS results. These wrinkles are largely in the armpits, abdomen, and chest area highlighted by the red rectangles. 

On the high level, the generalization of P-Net stems from the fact that it learns to predict the RB based on the garment hinge-graph GNN, the initial deformation, and the body motion. The initial deformation is influenced by the initial body shape. But this is somewhat counter-intuitive as wrinkle formation should be heavily influenced by the body deformation too, rather than the motion and the initial body shape only. However, in our method, the body deformation is handled by the E-Net. In other words, P-Net predicts the likely plasticity for the whole motion and E-Net adds elasticity caused by the body deformation. 

\begin{table}[tb]
    \centering
    \caption{Numerical comparison on unseen bodies. Metric are averages across all bodies.}
    \resizebox{\textwidth}{!}{
    \begin{tabular}{lccccccccc}
        \toprule
Base/Mtr & BED(rad)$\downarrow$ & PE(rad)$\downarrow$ & BE(rad)$\downarrow$ & MED($m$)$\downarrow$ & CD($m^2$)$\downarrow$ & BEN$\downarrow$ & COL($m$)$\downarrow$ \\
        \midrule
PBNS & 0.66252 & 0.14416 & 0.20414 & 0.05169 & 0.00078 & 0.03053 & \textbf{0.00626} \\
SNUG & 0.66125 & 0.14416 & 0.21759 & 0.02773 & 0.00040 & 0.03657 & 0.00773 \\
NCS & 0.74892 & 0.14416 & 0.19755 & 0.03444 & 0.00032 & 0.02778 & 0.00756 \\
HOOD & 0.69630 & 0.14416 & 0.21796 & 0.04612 & 0.00090 & 0.07248 & 0.00654 \\
SENC & 0.87741 & 0.14416 & 0.19979 & 0.03686 & 0.00066 & 0.05565 & 0.00652 \\
Ours & \textbf{0.46630} & \textbf{0.09690} & \textbf{0.19593} & \textbf{0.02402} & \textbf{0.00024} & \textbf{0.02252} & 0.00754 \\
        \bottomrule
    \end{tabular}
    }
    \label{tab:gen_bodies}
\end{table}

\cref{tab:gen_bodies} shows the quantitative results on unseen body shapes. Our method achieves the best performance across all evaluated metrics, demonstrating strong generalization beyond the body shapes used during training. For each metric, comparing with the best-performing baseline, our approach improves the performance by 29.48\% in BED, 32.78\% in PE, 0.82\% in BE, 13.38\% in MED, 25.00\% in CD, and 18.93\% in BEN, while maintaining competitive COL performance. In particular, we obtain the lowest BED, BE, and BEN, indicating that our predicted bending behavior and wrinkle distribution align most closely with the PBS under body-shape variation. 

Despite the overall good results, we still notice visual differences between our results and PBS. The primary limitation stems from the deformation pipeline: our method, similar to prior self-supervised simulators such as NCS and SNUG, relies on skinning-based garment attachment to the body. As a result, garment deformation is implicitly constrained to follow the body surface, leading to tight fitting behaviors. Although gravity is included as an energy term in these methods, gravity alone is insufficient to produce realistic loose draping. In the absence of explicit contact resolution and inequality-based collision constraints, the energy-based formulation cannot create stable garment–body separation. The garment is encouraged to minimize penetration, but it is not allowed to detach and re-drape freely under gravity. In contrast, PBS explicitly solves body–garment collisions and contact forces at each time step, enabling garments to be lifted, separated from the body, and naturally draping under gravity.

\subsection{Persistent Wrinkles on Unseen Garments}

\begin{figure*}[tb]
    \centering
    \includegraphics[width=\textwidth]{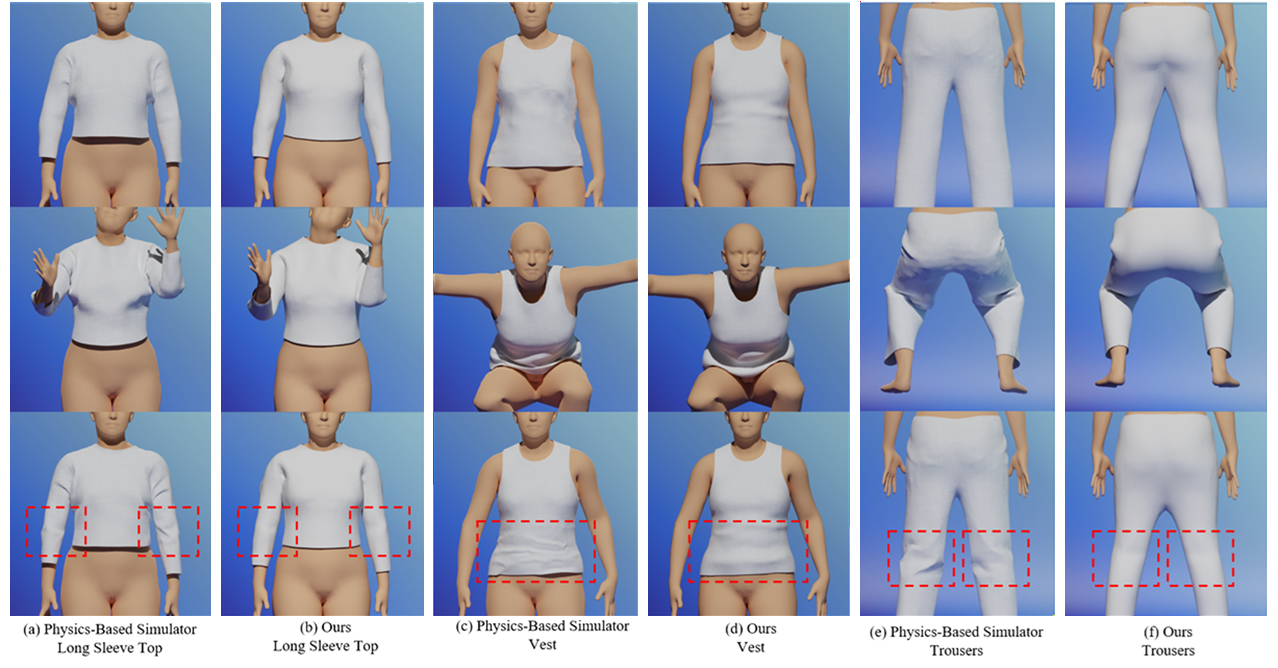}
    \caption{A comparison of wrinkle patterns across different garment types. (a), (c), and (e) present the PBS simulation results for the long-sleeve top, vest, and T-shirt, respectively, whereas (b), (d), and (f) show the corresponding predictions generated by our model.}
    \label{fig:diff_garment}
\end{figure*}

We further evaluate our method on unseen garment types, including long-sleeve top, vest, and trousers. Although P-Net is trained only on the t-shirt, it generalizes effectively to new garment geometries. As shown in \cref{fig:diff_garment}, our method produces wrinkle distributions (b, d, f) that closely resemble those generated by the PBS (a, c, e), demonstrating visually consistent persistent wrinkle formation across different garment structures.

The quantitative comparison in \cref{tab:gen_garments} further supports this observation. Our method achieves the best performance across all reported metrics. For each metric, comparing with the best-performing baseline, our approach improves the performance by 5.13\% in BED, 36.23\% in PE, 12.94\% in BE, 5.63\% in MED, 20.00\% in CD, and 12.65\% in BEN, without degrading COL performance. The reduced PE confirms that the learned plastic strain evolution transfers across mesh topologies, while the improvements in bending and geometric errors indicate more accurate wrinkle localization and overall deformation fidelity. These results suggest that modeling plasticity on an intrinsic hinge graph enables topology-aware generalization, allowing the simulator to adapt to diverse garment configurations without relying on garment-specific templates. 

\begin{table*}[tb]
    \centering
    \caption{Numerical comparison on unseen garments. Metrics are averaged across all testing garments.}
    \resizebox{\textwidth}{!}{
    \begin{tabular}{lccccccccccccc}
        \toprule
        Base/Mtr & BED(rad)$\downarrow$ & PE(rad)$\downarrow$ & BE(rad)$\downarrow$ & MED($m$)$\downarrow$ & CD($m^2$)$\downarrow$ & BEN$\downarrow$ & COL($m$)$\downarrow$ \\
        \midrule
PBNS
& 0.68964 & 0.17022 & 0.22727 & 0.03946 & 0.00061 & 0.04194 & \textbf{0.00022}\\
SNUG
& 0.55103 & 0.17022 & 0.20903 & 0.03447 & 0.00068 & 0.07346 & 0.00255\\
NCS
& 0.70959 & 0.17022 & 0.18726 & 0.03488 & 0.00060 & 0.03454 & 0.00241\\
HOOD
& 0.45466 & 0.17022 & 0.21034 & 0.05640 & 0.00102 & 0.15023 & 0.00353\\
SENC
& 0.44631 & 0.17022 & 0.20987 & 0.05788 & 0.00101 & 0.16064 & 0.00114\\
Ours
& \textbf{0.42340} & \textbf{0.10855} & \textbf{0.16303} & \textbf{0.03253} & \textbf{0.00048} & \textbf{0.03017} & 0.00237\\
        \bottomrule
    \end{tabular}
    }
    \label{tab:gen_garments}
\end{table*}

\subsection{Simulating User-specified Materials}

\begin{figure*}[tb]
    \centering
    \includegraphics[width=0.6\textwidth]{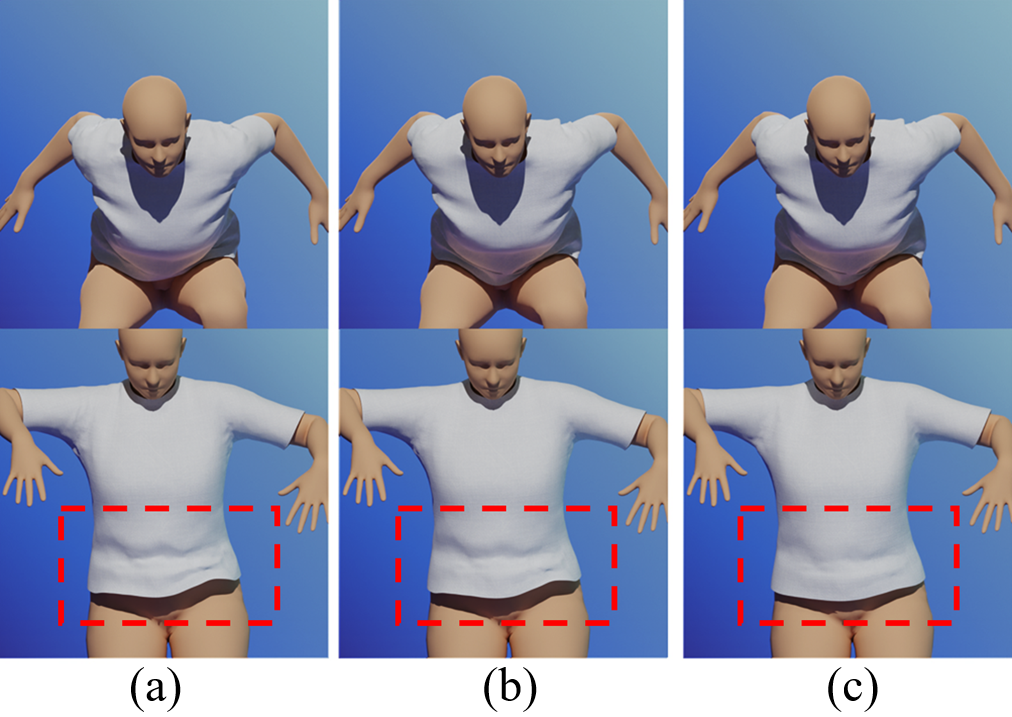}
    \caption{
Our simulator allows users to tweak the fabric plastic properties.
(a) Fabric tends to have wrinkles. By setting $\varepsilon_y$ to $30^\circ$, the simulated garment, like paper, is very likely to form wrinkles after a bending motion.
(b) Fabric is less likely to have wrinkles. After increasing $\varepsilon_y$ to $60^\circ$, the garment forms fewer wrinkles.
(c) Fabric is not likely to have wrinkles. As $\varepsilon_y$ is set to $90^\circ$, the garment becomes more wrinkle-resistant.
}
    \label{fig:diff_materials}
\end{figure*}

An important property of garment simulators is the ability to control material behavior. In physics-based methods, this is achieved by directly adjusting constitutive parameters, whereas existing self-supervised neural simulators mainly tune the weights of elastic energy terms. In contrast, our formulation explicitly models plasticity and introduces a physically meaningful yield threshold $\varepsilon_y$ in \cref{eq:delta_plastic}, which controls the onset of plastic deformation. A smaller $\varepsilon_y$ makes plastic accumulation easier, leading to more pronounced and persistent wrinkles. As shown in \cref{fig:diff_materials}, setting $\varepsilon_y$ to $30^\circ$, $60^\circ$, and $90^\circ$ progressively increases the resistance to wrinkle formation under the same motion. When $\varepsilon_y = 30^\circ$, plastic deformation is triggered early, resulting in clearly visible persistent folds. In contrast, when $\varepsilon_y = 90^\circ$, significantly larger bending is required before plastic accumulation occurs, and the garment behaves more elastically. Importantly, this qualitative change in wrinkle persistence cannot be reproduced by merely adjusting other parameters such as the bending stiffness $k_b$ or the plastic gain $k_p$, which only affect stiffness or update smoothness but do not alter the yield mechanism itself. Notably, even when $\varepsilon_y = 30^\circ$, the accumulated RB may exceed $30^\circ$, as the threshold controls the onset of plasticity rather than its upper bound.

\subsection{Ablation Study}
P-Net is the core component responsible for modeling the dynamics of the evolving RB. To justify our architectural design, we conduct ablation studies comparing different alternatives for spatial–temporal modeling of the internal plastic strain. Specifically, we evaluate GCN plus recurrent neural networks (RNN), long short-term memory (LSTM), Transformer, and a hinge-graph-based GNN with recurrent message passing. As shown in \cref{tab:ablation_study_spatial_temporal}, the Transformer achieves competitive accuracy compared to the proposed GNN, particularly on geometric metrics. However, it incurs substantially higher computational cost and significantly lower inference speed. In contrast, the hinge-graph GNN with message passing not only achieves the best overall performance on key metrics, but also requires dramatically less GPU memory and delivers an order-of-magnitude improvement in frames per second (FPS). These results indicate that hinge-level message passing is both more efficient and better aligned with the intrinsic mesh structure of plastic strain update. Therefore, we adopt the GNN with message passing as the default architecture for P-Net.

\begin{table}[tb] 
\centering 
\caption{Ablation Study. Alternative networks for learning the RB dynamics.} 
\resizebox{\textwidth}{!}{ 
\begin{tabular}{lccccccccc} 
\toprule

Base/Mtr & BED(rad)$\downarrow$ & PE(rad)$\downarrow$ & BE(rad)$\downarrow$ & MED(m)$\downarrow$ & CD($m^2$)$\downarrow$ & BEN$\downarrow$ & GPU(GB) $\downarrow$ & FPS$\uparrow$\\ 
\midrule 
RNN & 0.86971 & 0.09968 & \textbf{0.17048} & 0.03274 & 0.00043 & \textbf{0.02937} & 9.1 & 1.40\\ 

LSTM & 0.84272 & 0.07535 & 0.17149 & 0.03251 & 0.00042 & 0.02940 & 8.5 & 1.61\\ 

Transformer & 0.52095 & 0.05263 & 0.17482 & 0.02460 & 0.00029 & 0.03027 & 9.2 & 0.17\\ 

GNN & \textbf{0.28714} & \textbf{0.03206} & 0.17118 & \textbf{0.02355} & \textbf{0.00027} & 0.02999 & \textbf{0.6} & \textbf{9.43}\\ 

\bottomrule 
\end{tabular} } 
\label{tab:ablation_study_spatial_temporal} 
\end{table}

\begin{figure}[tb]
    \centering
    \includegraphics[width=\linewidth]{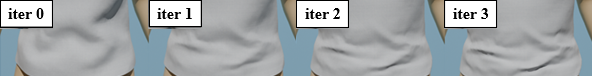}
    \caption{Convergence behavior under different alternating cycle settings.}
    \label{fig:iteration_test2}
\end{figure}

\begin{figure}[tb]
    \centering

    \begin{subfigure}[t]{0.49\linewidth}
        \centering
        \includegraphics[width=\linewidth]{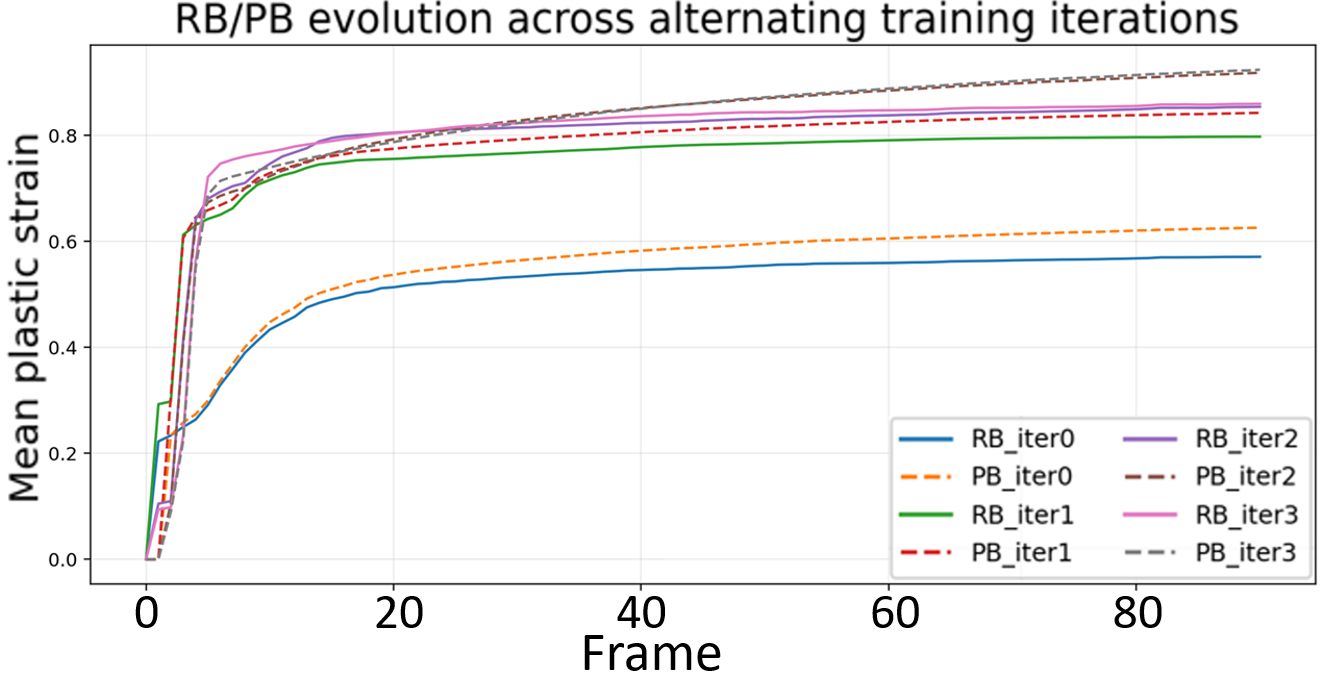}
        \caption{}
        \label{fig:iteration_curve}
    \end{subfigure}
    \hfill
    \begin{subfigure}[t]{0.49\linewidth}
        \centering
        \includegraphics[width=\linewidth]{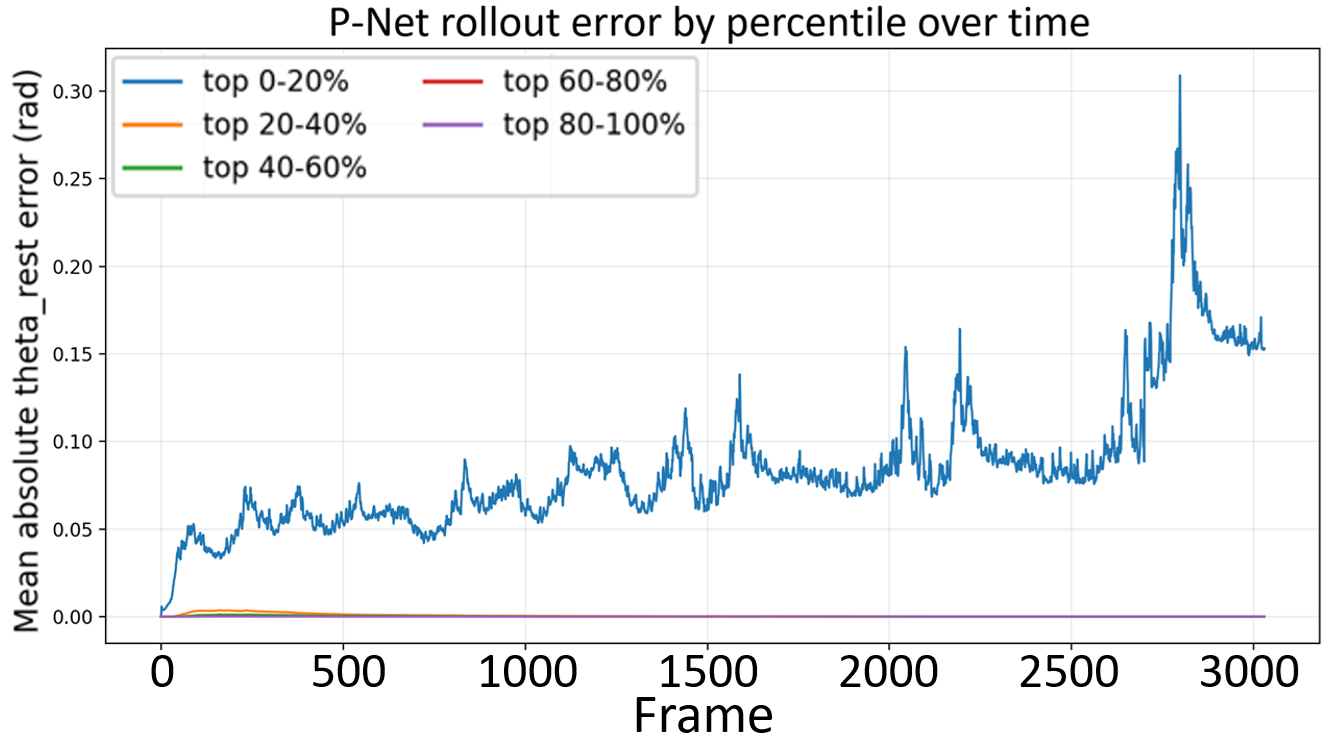}
        \caption{}
        \label{fig:long_term_error}
    \end{subfigure}

    \caption{
    Training dynamics and long-term stability of our method.
    (a) Rest Bending (RB) and Predicted Rest Bending (PB) evolution across alternating training iterations.
    (b) Long-term autoregressive error over rollout frames.
    }
    \label{fig:training_analysis}
    \vspace{-1em}
\end{figure}

Our model is trained using 4 alternating training cycles. As shown in \cref{fig:iteration_curve}, curriculum learning gradually evolves the plastic state from near-elastic behavior toward a converged elasto-plastic solution, and from \cref{fig:iteration_test2}, we observe that increasing alternating cycles beyond four provides small improvement. 

We additionally tested 10 randomly selected long motion sequences. These sequences were randomly sampled from the evaluation set. At each frame, samples are ranked by error and divided into five percentile groups (top 0–20\% to top 80–100\%). As shown in \cref{fig:long_term_error}, the average top-20\% error remains around 0.05 for the first 700 frames and around 0.1 up to approximately 2500 frames, indicating stable long-term generation. Significant drift only appears after extremely long horizons ($\sim$3000 frames). Most edges maintain low prediction error throughout the rollout.

\section{Discussions and Conclusions}

We have proposed the first self-supervised neural garment simulator that is capable of simulating persistent wrinkles. We model garments as an elasto-plastic material and design two networks, E-Net and P-Net, correspondingly. P-Net predicts the RB, enabling E-Net to keep the garment in non-flat state and generating persistent wrinkles. We have also proposed a novel curriculum learning method to address the difficulty in learning caused by the moving energy minimization problem. This curriculum learning strategy is physics-inspired and achieves convergence with a few cycles. We have demonstrated that our simulator can more closely mimic physics-based simulators in persistent wrinkle formation in unseen motions, bodies, and garments.

Our simulator has limitations. Our model cannot simulate wrinkles being flattened by stretching. Moreover, we do not evaluate on loose or multi-layered garments, which are orthogonal to our research. In future work, we will address these limitations. Moreover, we will incorporate other existing SSL models as instantiation of our E-Net, to make our model a plug-and-play component.

\section{Acknowledge}
This work was partially funded by the HKSAR ITC under ITSP-Platform grants (Ref: ITS/335/23FP, ITS/469/24FP).

\bibliographystyle{splncs04}
\bibliography{main}

% ---- Arxiv Supplementary Material ----
\appendix
\title{Self-supervised Garment Dynamics with Persistent Wrinkles Supplementary Material} 

\titlerunning{Self-supervised Garment Dynamics with Persistent Wrinkles}

\author{
Xiaoyuan Yang\inst{1}\orcidlink{0009-0001-1379-8918} \and
Deshan Gong\inst{3}\orcidlink{0009-0002-2516-9542} \and
Taku Komura\inst{3}\orcidlink{0000-0002-2729-5860} \and
He Wang\thanks{corresponding author, he\_wang@ucl.ac.uk}\inst{1,2}\orcidlink{0000-0002-2281-5679}
}

\authorrunning{X. Yang et al.}

\institute{
$^1$University of Leeds,
$^2$University College London,
$^3$The University of Hong Kong
}

\maketitle

We refer the reader to \textbf{our video} for additional visual results. Our code is publicly available at \textit{https://github.com/realcrane/EPNet}

\section{Additional Experimental Results}
\subsection{Full metrics}
In addition to the metrics reported in the main paper, namely Bending Error Distribution (BED), Plastic Error (PE), Bending Error (BE), Mean Euclidean Distance (MED), Chamfer Distance (CD), Bending Energy (BEN), and Collision Error(COL), we further include Stretching Energy (STE), Shearing Energy (SHE), Gravity Energy (GE), and Inertia Energy (IE). Although these metrics are not directly related to wrinkle formation, we include them for completeness in the comparison with baselines. Lower values indicate better performance for all metrics. Next, we give the details of all metrics we employ.

BED is the per-edge $\mathrm{L1}$ error computed over a group of edges. This group is selected using Segmented Maximum Analysis based on the maximum Plastic Bending (PB) observed across all frames and motions. We select the top 10\% edges and compute BED by:
\begin{equation}
    \mbox{BED} = \frac{1}{T D^\prime} \sum^T_{t} \sum^{D^\prime}_{d}  \| \varepsilon_d^{(t)} - \hat{\varepsilon}_d^{(t)} \|
    \label{eq:app_mtr_bed}
\end{equation}
where $T$ is the number of frames and $D^\prime$ denotes the number of selected edges. $\varepsilon$ and $\hat{\varepsilon}$ denote the predicted strain from the self-supervised simulator and the ground-truth strain from the physics-based simulator (PBS), respectively.

BE is the per-edge $\mathrm{L1}$ error of the predicted $\varepsilon$ from P-Net.
\begin{equation}
    \mbox{BE} = \frac{1}{TD} \sum^T_{t} \sum^D_{d} \| \varepsilon_d^{(t)} - \hat{\varepsilon}_d^{(t)} \|
    \label{eq:app_mtr_be}
\end{equation}
where $D$ is the number of all garment mesh edges.

PE is the per-edge $\mathrm{L1}$ error of the final $\varepsilon_p$ predicted by P-Net. Note that PBNS, SNUG, NCS, HOOD, and SENC do not model PB, so their PB is effectively zero.
\begin{equation}
    \mbox{PE} = \frac{1}{TD} \sum^T_{t} \sum^D_{d} \| \varepsilon_{p,d}^{(t)} - \hat{\varepsilon}_{p,d}^{(t)} \|
    \label{eq:app_mtr_pe}
\end{equation}
where $\varepsilon_{p,*}$ and $\hat{\varepsilon}_{p,*}$ are the predicted plastic strain from self-supervised simulator and the ground-truth plastic strain from PBS, respectively.

MED measures the average Euclidean distance between the vertex positions of the predicted garment mesh, $\mathbf{x}$, and the ground-truth mesh from PBS, $\hat{\mathbf{x}}$:
\begin{equation}
    \mbox{MED}(\mathbf{x}, \hat{\mathbf{x}})
= \frac{1}{TN} \sum_{t}^T \sum_{i}^N \| \mathbf{x}_i^{(t)} - \hat{\mathbf{x}}_i^{(t)} \|
    \label{eq:app_mtr_med}
\end{equation}
where $\mathbf{x}_i^{(t)} \in \mathbb{R}^3$ is the position of the $i$th garment mesh vertex at the $t$-th frame, and N is the number of vertices

CD measures the Chamfer Distance between the garment mesh from self-supervised cloth simulator, $\mathbf{x}$, and the mesh from PBS, $\hat{\mathbf{x}}$: 
\begin{align}
\mbox{CD}(\mathbf{x}, \hat{\mathbf{x}}) &= \frac{1}{TN} \sum_{t}^T \sum_{i}^N \min_j \| \mathbf{x}_i^{(t)} - \hat{\mathbf{x}}_j^{(t)} \|_2^2 \notag \\
&+ \frac{1}{TN} \sum_{t}^T \sum_{i}^N \min_j \| \mathbf{x}_j^{(t)} - \hat{\mathbf{x}}_i^{(t)} \|_2^2
    \label{eq:app_mtr_cd}
\end{align}
where $j \in \{1, \dots, N\}$.

BEN measures garment bending energy:
\begin{equation}
    \mbox{BEN} = \frac{1}{TD} \sum^T_{t} \sum^D_{d} k_b(\varepsilon_d^{(t)} - \varepsilon_{p,d}^{(t)})^2
    \label{eq:app_mtr_ben}
\end{equation}
where $k_b$ is the bending stiffness. 

COL measures the penetration depth between the garment vertices and the body mesh:
\begin{align}
\mbox{COL} &= \frac{1}{TN} \sum_{t}^T \sum^{N}_{i} \|\min(f_{dis}(\mathbf{x}^{(t)}_{i}), 0)\|
    \label{eq:app_mtr_COL}
\end{align}

where $f_{dis}(\mathbf{x}_t)$ measures the signed distance between a garment vertex position and the body mesh. 

STE, SHE, GE, and IE are computed following the formulation described in \cref{sec:app_lossTerms}.

\subsection{Full Details of Comparisons} 

\begin{figure*}[tb]
    \centering
    \includegraphics[width=\textwidth]{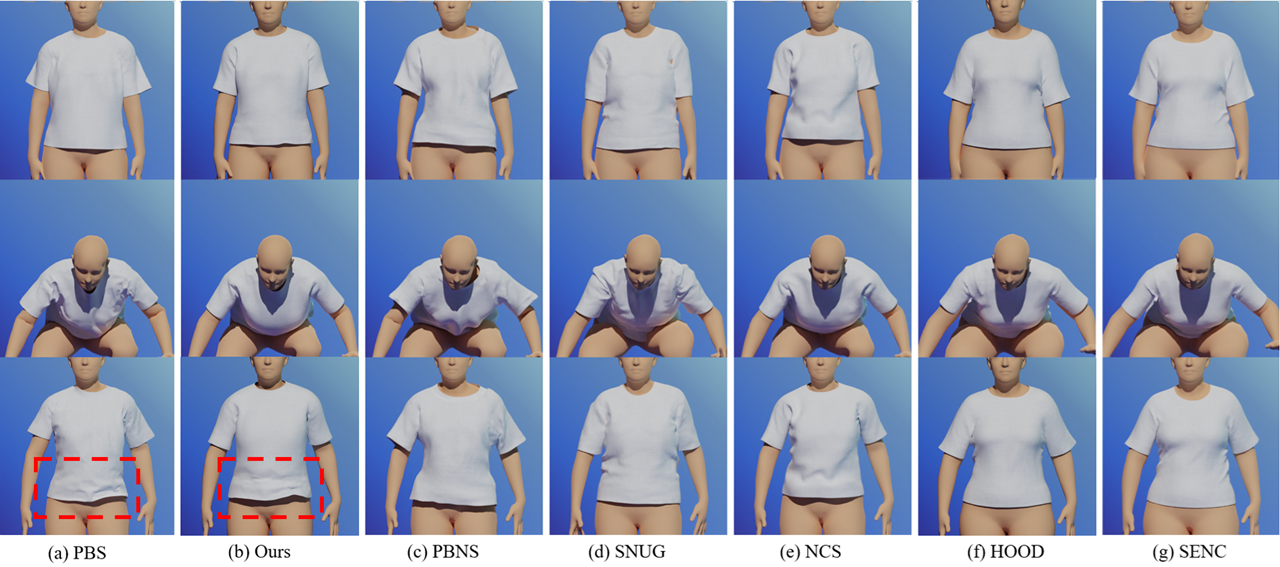}
    \caption{Compare the persistent wrinkles formed by a waist bending motion. By modeling cloth as elasto-plastic material, Physics-based Simulator (a) can simulate the persistent wrinkles in the belly area of the t-shirt. Our simulator is capable of closely mimicking the wrinkles (b). However, the baseline method either cannot simulate persistent wrinkles after the waist bending motion (c, d, f, g) or generate folds in the different areas even before bending happened (e). }
    \label{fig:app_diff_motion}
\end{figure*}

\begin{figure*}[tb]
    \centering
    \includegraphics[width=\textwidth]{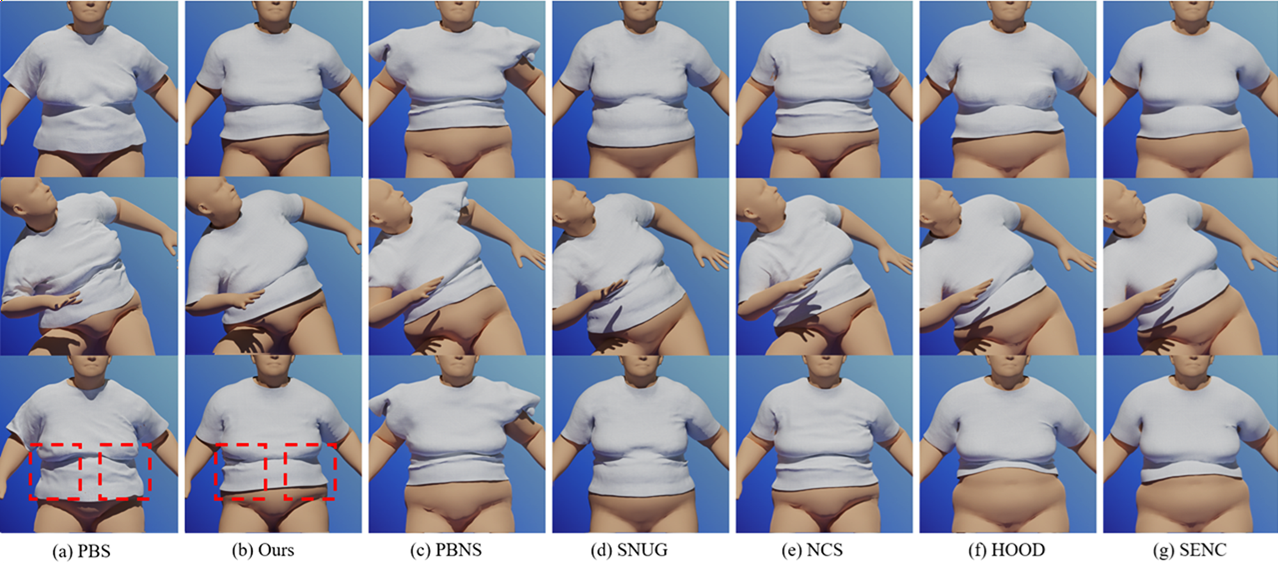}
    \caption{Compare the persistent wrinkles formed on a t-shirt worn by a obese body. The results demonstrate our simulator (b) can constantly simulate persistent wrinkles which are close to PBS (a) even if the body shape is varied. Whereas the garments simulated by the baseline methods (c-g) exhibit wrinkle behaviors that differ from those in the PBS.}
    \label{fig:app_diff_body_tshirt}
\end{figure*}

Visually, \cref{fig:app_diff_motion,fig:app_diff_body_tshirt} demonstrates that our simulator captures persistent wrinkles similar to those produced by PBS across different motions and body shapes. In contrast, the baseline methods cannot reproduce the persistent wrinkles caused by material plasticity. Instead, the folds in their results are mainly induced by body–garment collisions and appear visually different from the wrinkles produced by PBS.
Quantitatively, we also compare the metrics of our simulator and the baseline methods in \cref{tab:app_gen_motions,tab:app_gen_bodies}. Our method does not achieve the best scores in GE and IE, but the differences from the best-performing methods are relatively small. Overall, we observe that GE and IE are similar across most methods. GE mainly reflects differences in vertex height, while IE measures vertex acceleration. These quantities are not directly influenced by plastic deformation.
Based on these observations, as well as the results presented later, we speculate that the variations in GE and IE may be partly caused by training randomness, such as network initialization. We report GE and IE following prior works for completeness; however, these metrics do not measure the quality of wrinkle formation and are therefore not directly related to the main contribution of our work.


\begin{table*}[tb]
    \centering
    \caption{
    Numerical comparison on unseen motions. Metrics are averaged across all testing motions (the lower the better).
    }
    \resizebox{\textwidth}{!}{
    \begin{tabular}{lccccccccccccc}
        \toprule
        Base/Mtr & BED(rad)$\downarrow$ & PE(rad)$\downarrow$ & BE(rad)$\downarrow$ & MED($m$)$\downarrow$ & CD($m^2$)$\downarrow$ & BEN$\downarrow$ & COL($m$)$\downarrow$ & STE$\downarrow$ & SHE$\downarrow$ & GE$\downarrow$ & IE$\downarrow$ \\
        \midrule
PBNS & 0.60223 & 0.12506 & 0.18733 & 0.04038 & 0.00054 & 0.04565 & \textbf{0.00095} & 0.11225 & 0.17735 & \textbf{0.19897} & 0.00185 \\
SNUG & 0.64567 & 0.12506 & 0.20327 & 0.03191 & 0.00061 & 0.05976 & 0.00136 & 0.05132 & 0.05564 & 0.21234 & 0.00167 \\
NCS  & 0.76556 & 0.12506 & 0.17611 & 0.03210 & 0.00040 & 0.03412 &  0.00129 & 0.00419 & 0.01866 & 0.21938 & 0.00182 \\
HOOD  & 0.53553 & 0.12506 & 0.20948 & 0.06081
 & 0.00118 & 0.13150 & 0.00210 & 0.08521 & 0.17040 & 0.23601 & 0.00199 \\
SENC  & 0.52529 & 0.12506 & 0.19785 & 0.05486 & 0.00074 & 0.13657 &  0.00160 & 0.05006 & 0.07686 & 0.22472 & \textbf{0.00158} \\
Ours & \textbf{0.28714} & \textbf{0.03206} & \textbf{0.17118} & \textbf{0.02355} & \textbf{0.00027} & \textbf{0.02999} & 0.00129 & \textbf{0.00261} & \textbf{0.01214} & 0.21112 & 0.00180 \\
        \bottomrule
    \end{tabular}
    }
    \label{tab:app_gen_motions}
\end{table*}
\begin{table*}[tb]
    \centering
    \caption{Numerical comparison on unseen bodies. Metrics are averaged across all testing bodies (the lower the better).}
    \resizebox{\textwidth}{!}{
    \begin{tabular}{lcccccccccccc}
        \toprule
        Base/Mtr & BED(rad)$\downarrow$ & PE(rad)$\downarrow$ & BE(rad)$\downarrow$ & MED($m$)$\downarrow$ & CD($m^2$)$\downarrow$  & BEN$\downarrow$ & COL($m$)$\downarrow$ & STE$\downarrow$ & SHE$\downarrow$ & GE$\downarrow$ & IE$\downarrow$ \\
        \midrule
PBNS & 0.66252 & 0.14416 & 0.20414 & 0.05170 & 0.00078 & 0.03053 & \textbf{0.00626} & 0.10453 & 0.10768 & 0.23952 & 0.00121 \\
SNUG & 0.66125 & 0.14416 & 0.21759 & 0.02773 & 0.00040 & 0.03657 &  0.00773 & 0.35830 & 1.09185 & \textbf{0.23531} & 0.00142 \\
NCS & 0.74892 & 0.14416 & 0.19755 & 0.03444 & 0.00032 & 0.02778 & 0.00756 & 0.02245 & 0.05507 & 0.24299 & 0.00112 \\
HOOD  & 0.69630 & 0.14416 & 0.21796 & 0.04612 & 0.00090 
& 0.07248 &  0.00654 & 0.09127 & 0.10900 & 0.24312 & 0.00156 \\
SENC  & 0.87741 & 0.14416 & 0.19979 & 0.03686 & 0.00066
& 0.05565 & 0.00652 & 0.07168 & 0.05767 & 0.23591 & 0.00122 \\
Ours & \textbf{0.46630} & \textbf{0.09690} & \textbf{0.19593} & \textbf{0.02402} & \textbf{0.00024} & \textbf{0.02252} & 0.00754 & \textbf{0.02027} & \textbf{0.04246} & 0.23751 & \textbf{0.00112} \\
        \bottomrule
    \end{tabular}
    }
    \label{tab:app_gen_bodies}
\end{table*}

\begin{figure*}[tb]
    \centering
    \includegraphics[width=\textwidth]{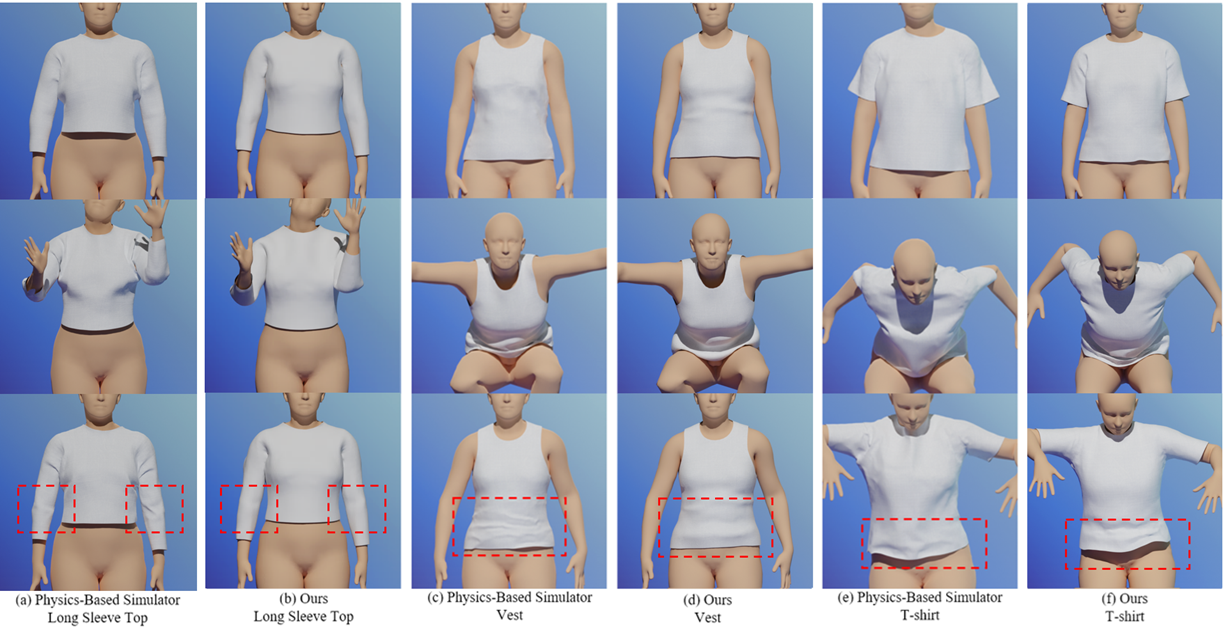}
    \caption{A comparison of wrinkle patterns across different garment types. (a), (c), and (e) present the PBS simulation results for the long-sleeve top, vest, and T-shirt, respectively, whereas (b), (d), and (f) show the corresponding predictions generated by our model.}
    \label{fig:app_diff_garment}
\end{figure*}

\begin{table*}[tb]
    \centering
    \caption{Numerical comparison on unseen garments. Metrics are averaged across all testing garments (the lower the better).}
    \resizebox{\textwidth}{!}{
    \begin{tabular}{lccccccccccccc}
        \toprule
        Base/Mtr & BED(rad)$\downarrow$ & PE(rad)$\downarrow$ & BE(rad)$\downarrow$ & MED($m$)$\downarrow$ & CD($m^2$)$\downarrow$ & BEN$\downarrow$ & COL($m$)$\downarrow$ & STE$\downarrow$ & SHE$\downarrow$ & GE$\downarrow$ & IE$\downarrow$ \\
        \midrule
PBNS
& 0.68964 & 0.17022 & 0.22727 & 0.03946 & 0.00061 & 0.04194 & \textbf{0.00022} & 0.04795 & 0.05617 & \textbf{0.14884} & 0.00278 \\

SNUG
& 0.55103 & 0.17022 & 0.20903 & 0.03447 & 0.00068 & 0.07346 & 0.00255 & 0.06346 & 0.07084 & 0.15633 & 0.00247 \\

NCS
& 0.70959 & 0.17022 & 0.18726 & 0.03488 & 0.00060 & 0.03454 & 0.00241 & 0.00199 & 0.00958 & 0.15657 & \textbf{0.00230} \\

HOOD
& 0.45466 & 0.17022 & 0.21034 & 0.05640 & 0.00102 & 0.15023 & 0.00353 & 0.08336 & 0.13087 & 0.16539 & 0.00292 \\

SENC
& 0.44631 & 0.17022 & 0.20987 & 0.05788 & 0.00101 & 0.16064 & 0.00114 & 0.07387 & 0.09422 & 0.16303 & 0.00241 \\

Ours
& \textbf{0.42340} & \textbf{0.10855} & \textbf{0.16303} & \textbf{0.03253} & \textbf{0.00048} & \textbf{0.03137} & 0.00237 & \textbf{0.00144} & \textbf{0.00674} & 0.15265 & 0.00239 \\

        \bottomrule
    \end{tabular}
    }
    \label{tab:app_gen_garments}
\end{table*}

\begin{figure*}[tb]
    \centering
    \includegraphics[width=\textwidth]{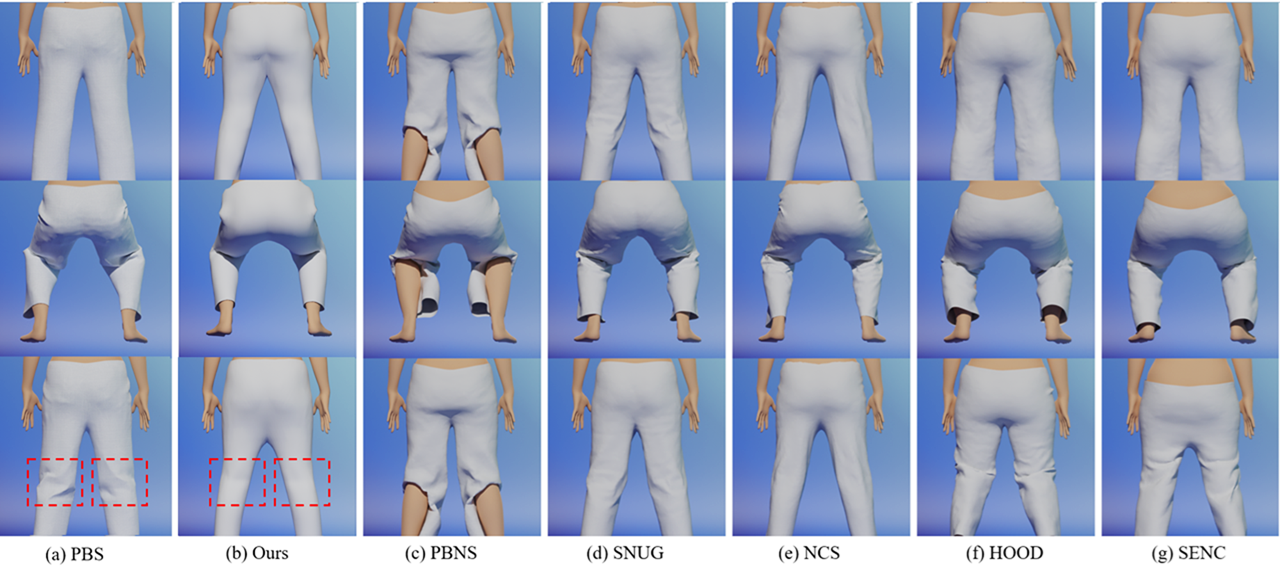}
    \caption{Compare the persistent wrinkles formed on a pair of trousers caused by squatting. Our simulator (b) can closely simulate the persistent wrinkles as the PBS (a). On the contrary, the trousers simulated by SNUG (d) and NCS (e) are nearly flat as the garment is modeled as elastic material. The folds in (f) and (g), i.e., by HOOD and SENC, are caused by body-garment collision, so the position of the folds are different from the persistent wrinkles simulated by PBS. Finally, PBNS (c) encounters severe penetration errors even if we use their official trained models. }
    \label{fig:app_diff_motion_pants}
\end{figure*}

\begin{figure*}[tb]
    \centering
    \includegraphics[width=\textwidth]{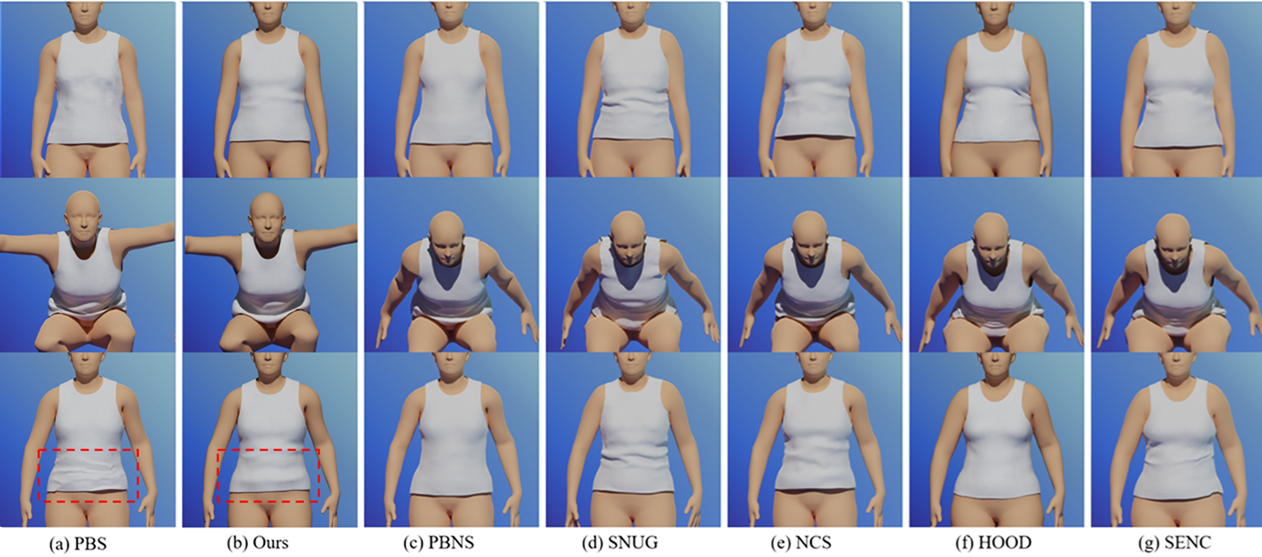}
    \caption{Compare the persistent wrinkles formed on a vest caused by a bending waist motion. The belly area of the vest is bended greatly due to the motion (middle row). Only our simulator (b) can closely simulate the persistent wrinkles as PBS (a) resulted from this motion. However, the vest simulated by PBNS (c) has no wrinkles. Although the other baseline methods (d-g) simulate folds, but those folds are incurred by the body-garment interactions and are formed on the areas where are different from PBS. }
    \label{fig:app_diff_garment_tank}
\end{figure*}

We further demonstrate the generalization ability of P-Net on unseen garments. Apart from body shapes, varying garment types are also supported, \ie, our P-Net can operate on garments that are unseen during training. We simulate trousers, a vest, and a long-sleeve top.
As shown in \cref{fig:app_diff_garment,fig:app_diff_motion_pants,fig:app_diff_garment_tank}, only our simulator reproduces the persistent wrinkles caused by motion. In \cref{fig:app_diff_garment}, even though the model is trained only on the t-shirt (f), our simulator is capable of handling the long-sleeve top and vest (b, d), and the simulated wrinkles conform well to those from PBS (a, c).
Thanks to the message passing design, our model can simulate wrinkles around the elbow and popliteal areas caused by arm and leg bending, which are not observed in the t-shirt training data (\cref{fig:app_diff_garment,fig:app_diff_motion_pants}). The results show that our simulator can consistently reproduce physically plausible wrinkles similar to those produced by PBS across diverse garments, including trousers, long-sleeve tops, and vest.
However, garments simulated by the baseline methods either remain overly flat or accumulate folds that differ from those produced by PBS. For example, \cref{fig:app_diff_motion_pants} (e) and \cref{fig:app_diff_garment_tank} (c) form no wrinkles after the garment undergoes deformation. On the other hand, the folds formed in \cref{fig:app_diff_motion_pants} (d, f, g) and \cref{fig:app_diff_garment_tank} (d–g) are caused by garment–body collisions and appear at locations different from those in PBS.
\cref{tab:app_gen_garments} further quantitatively highlights that our simulator more closely mimics PBS simulations than the baseline methods.

\subsection{Ablation Study}
In addition to the ablation results presented in the main paper, we provide the full details and discussion here. The ablation study mainly evaluates the spatial and temporal aspects of our material model.

\begin{figure}[tb]
    \centering
    \includegraphics[width=0.8\linewidth]{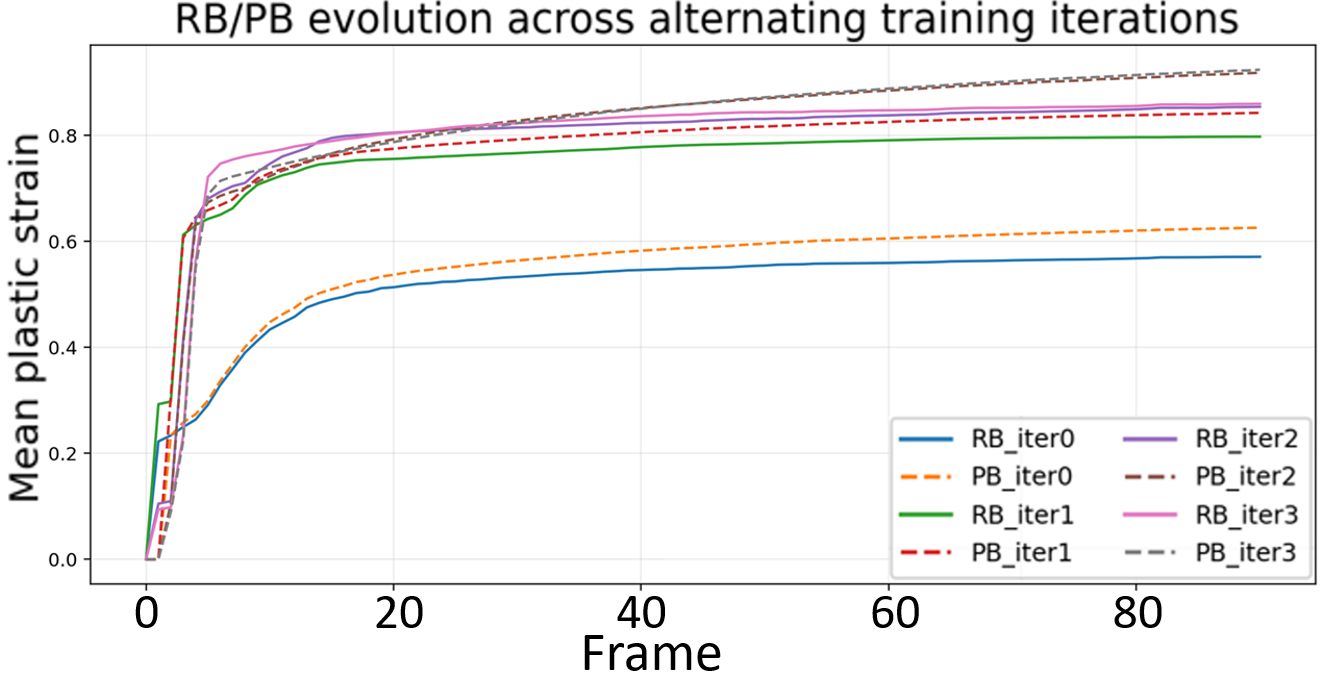}
    \caption{RB/PB evolution across alternating iterations.}
    \label{fig:app_iteration_curve}
\end{figure}

\noindent\textbf{The Necessity of P-Net and Curriculum Learning}
In inference, the core challenge of our SSL method is how to make E-Net predict plasticity-aware wrinkles, which requires the unknown plastic state as input. If E-Net were trained using an arbitrary plastic state, applying \cref{eq:app_epsilon_plastic} to the E-Net output would produce another plastic-state update. However, this updated state would still not be the converged one, as shown in \cref{fig:app_iteration_curve}. Feeding this updated state back into E-Net would further change the deformation prediction, which in turn would produce another plastic-state update. Consequently, \cref{eq:app_epsilon_plastic} and E-Net would need to be iteratively alternated until convergence. However, during training, the converged plastic state is unknown and no supervision exists for it. Therefore, P-Net is introduced to predict the evolving plastic state during rollout.

Furthermore, directly optimizing E-Net and P-Net often leads to unstable convergence and physically implausible plastic accumulation. Curriculum learning gradually evolves the plastic state from near-elastic behavior toward a converged elasto-plastic solution. As shown in \cref{fig:app_iteration_curve}, RB and PB play different roles: E-Net uses the evolving bending target, while P-Net predicts the plastic state from motion history. Thus, even if the converged RB approaches accumulated plastic bending, P-Net is not a trivial identity mapping.

\noindent\textbf{Spatial Bending Correlation}
Another important design choice in P-Net is the number of message-passing steps performed on the hinge graph. In our GNN architecture, a single message-passing layer allows information to propagate between a hinge and its 1-ring neighboring hinges. Stacking $n$ such layers enables the network to capture correlations within an $n$-ring neighborhood.
In principle, bending deformations in garments may exhibit long-range correlations. Existing neural simulators~\cite{santesteban2022snug} implicitly model global correlations by processing all edges together in a shared network. However, in our case, P-Net is designed specifically to model plastic bending. According to physics-based simulators~\cite{narain2013folding, gong2025cloth}, plastic deformation is primarily a local phenomenon determined by nearby geometric and material interactions. Therefore, it is more appropriate to model plastic evolution using a local neighborhood on the hinge graph.
Nevertheless, it is unclear how large this neighborhood should be in practice. To investigate this, we evaluate P-Net variants with 1, 2, and 3 message-passing steps, corresponding to 1-ring, 2-ring, and 3-ring hinge neighborhoods.

\begin{figure}[tb]
    \centering
    \includegraphics[width=\linewidth]{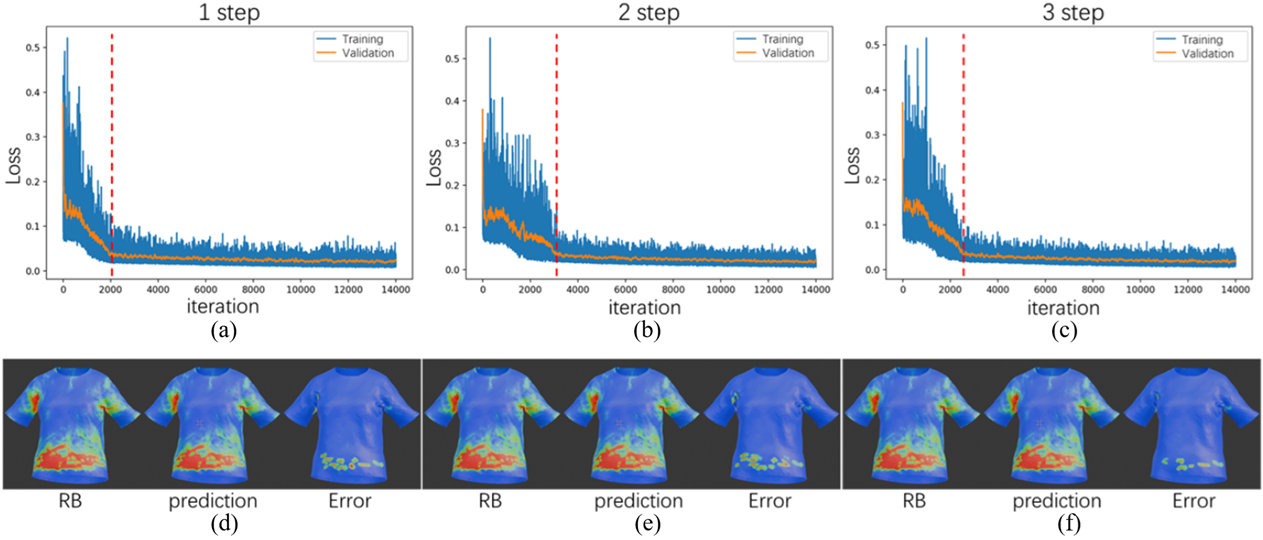}
    \caption{Ablation study on the number of message-passing steps in P-Net. 
The three columns correspond to 1-step, 2-step, and 3-step message passing.
(a–c) Training curves of P-Net. The blue curves represent the training loss and the orange curves represent the validation loss. 
The red dashed line indicates the approximate convergence point during training.
(d–f) Qualitative comparison of predicted rest bending (RB). 
For each example, the three garments show the target RB, the RB predicted by P-Net, and the corresponding error. 
The heatmaps visualize the hinge dihedral angles mapped from edges to vertices for visualization. 
The color scale ranges from blue (0) to green and red as the value increases, with the maximum value corresponding to $\pi$.}
    \label{fig:app_ablation_spatial}
\end{figure}

The quantitative results are reported in \cref{tab:app_ablation_study_spatial} and \cref{fig:app_ablation_spatial}. Increasing the neighborhood size improves P-Net performance by capturing broader spatial correlations. With 1-step message passing, the model converges earlier during training but results in higher mean and maximum RB errors. Using 2-step message passing reduces the mean RB error but leads to a higher maximum RB error and requires more training iterations to converge. With 3-step message passing, the model achieves both faster convergence and the best performance in terms of both mean and maximum RB errors.

Overall, these results suggest that using three message-passing steps provides a good balance between modeling spatial correlations and maintaining stable optimization.

\begin{table*}[tb]
    \centering
    \caption{Ablation Study on Spatial Bending Correlation}
    \begin{tabular}{lcc}
        \toprule
        Base/Mtr & Mean RB Error(rad)$\downarrow$ & Max RB Error(rad)$\downarrow$ \\
        \midrule
1-step & 0.0281575 & 2.9959853 \\
2-step & 0.0176077 & 3.0522114 \\
3-step (Ours) & \textbf{0.017319} & \textbf{2.975552} \\
        \bottomrule
    \end{tabular}
    \label{tab:app_ablation_study_spatial}
\end{table*}

\noindent\textbf{Temporal Dynamics of Plastic Bending}
In physics, (perfect) plasticity can be viewed as a dynamical system whose state is the rest bending (RB) angle. The evolution of RB depends on two factors: the current RB and the current total strain. However, to predict RB for all frames in one pass during inference, the trained network must capture temporal dependencies to reflect causality. 
If there are two RB changes in frames $t_1$ and $t_2$, and no other RB changes occur between them, we expect $t_2 > t_1$ when $RB^{t_2} > RB^{t_1}$. Moreover, $|t_2 - t_1|$ can be arbitrarily large, implying that the temporal dependence may span long time horizons. To investigate this, we also tested simpler temporal models including Recurrent Neural Networks (RNNs), Long Short-Term Memory networks (LSTMs), and Transformers.

The RNN layer has 256 hidden units, following the standard Elman recurrent formulation with a tanh state update and no gating mechanisms. The RNN produces a full sequence of hidden states, each of which is mapped to a scalar RB prediction through a linear projection layer. This block provides the most minimal recurrent baseline, without gating, attention, or graph-aware operations.

The LSTM extends this baseline by replacing the RNN layer with a single LSTM layer of 256 hidden units, following the standard LSTM formulation with sigmoid gating and a tanh state update. As with the RNN variant, the LSTM outputs a sequence of hidden states, which are projected to per-frame scalar values through a linear layer. Compared with the RNN, the gating mechanism allows the LSTM to better capture long-range temporal dependencies.

The Transformer adopts an encoder–decoder architecture. On the encoder side, we linearly project the per-node features to a 32-dimensional embedding, add sinusoidal positional encodings along the temporal axis, and process the sequence with a single pre-norm Transformer encoder block comprising 8-head self-attention and a two-layer feed-forward network with 128 hidden units and GELU activation. The encoder output is reshaped and flattened over the graph nodes to form a per-frame context vector. 
On the decoder side, the input sequence is slightly perturbed with Gaussian noise, prepended with a learnable Begin-of-Sequence token, projected to 32 dimensions, and enriched with sinusoidal positional encodings before being processed by a single pre-norm Transformer decoder block with causal 8-head self-attention and cross-attention to the encoder context. Finally, the decoder output is normalized and mapped by a three-layer MLP (32 → 128 → 128 → 128 → 1) to produce an RB value for each time step.

\begin{figure}[tb]
    \centering
    \includegraphics[width=\linewidth]{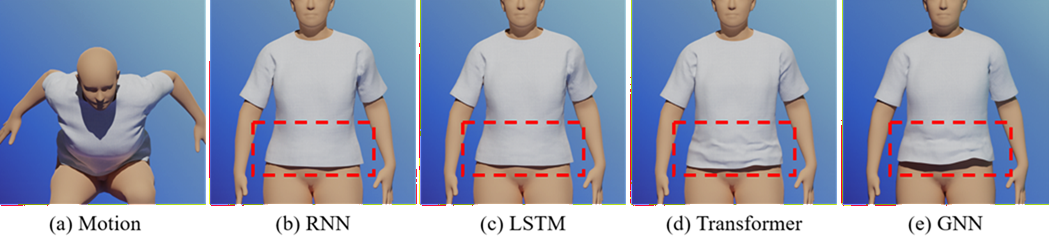}
    \caption{Visual comparison of different models for plastic bending dynamics. Compared with RNN (b) and LSTM (c), Transformer (d) and our message-passing GNN (e) can simulate more obvious wrinkles.}
    \label{fig:app_ablation_temporal}
\end{figure}

We first present the quantitative comparison in \cref{tab:app_ablation_study_temporal}. LSTM performs the worst among all methods. RNN occasionally achieves slightly better results in BE and BEN; however, this improvement appears to come from overall under-predicting the plastic bending. This behavior can also be observed in \cref{fig:app_ablation_temporal}, where both RNN and LSTM produce overly smooth garment deformations.
Both the Transformer and our message-passing GNN are able to generate persistent wrinkles. However, as shown in \cref{tab:app_ablation_study_temporal}, the Transformer requires significantly more GPU memory and substantially longer inference time. In contrast, the message-passing GNN achieves a better balance between modeling temporal correlations and computational efficiency.

\begin{table*}[tb]
    \centering
    \caption{Ablation Study on Plastic Bending Dynamics}
    \resizebox{\textwidth}{!}{
    \begin{tabular}{lcccccccccccc}
        \toprule
        Base/Mtr & BED(rad) $\downarrow$  & PE(rad) $\downarrow$  & BE(rad) $\downarrow$  & MED($m$) $\downarrow$  & CD($m^2$) $\downarrow$  & BEN $\downarrow$  & STE $\downarrow$  & SHE $\downarrow$  & GE $\downarrow$  $\downarrow$  & IE $\downarrow$  & GPU(GB) $\downarrow$ & FPS$\uparrow$\\
        \midrule
RNN         
& 0.86971 & 0.09968 & \textbf{0.17048} & 0.03274 & 0.00043 & \textbf{0.02937} & 0.03937 & 0.03367 & 0.21221 & 0.00184 & 9.1 & 1.40\\

LSTM        
& 0.84272 & 0.07535 & 0.17149 & 0.03251 & 0.00042 & 0.02940 & 0.03833 & 0.03299 & 0.21235 & 0.00184 & 8.5 & 1.61\\

Transformer  
& 0.52095 & 0.05263 & 0.17482 & 0.02460 & 0.00029 & 0.03027 & \textbf{0.00252} & \textbf{0.01177} & 0.21217 & 0.00198 & 9.2 & 0.17\\

GNN (Ours)
& \textbf{0.28714} & \textbf{0.03206} & 0.17118 & \textbf{0.02355} & \textbf{0.00027} & 0.02999 & 0.00261 & 0.01214 & \textbf{0.21112} & \textbf{0.00180} & \textbf{0.6} & \textbf{9.43}\\
        \bottomrule
    \end{tabular}
    }
    \label{tab:app_ablation_study_temporal}
\end{table*}

% \paragraph{Elastic Net Part} Our P-Net can be appended to other elastic self-supervised cloth simulators to simulate elasto-plastic garments. As shown in the \DS{Figure}. As shown in the~\cref{tab:other_elastic}

% \begin{table*}[tb]
%     \centering
%     \caption{Elastic Net +}
%     \resizebox{\textwidth}{!}{
%     \begin{tabular}{lcccccccccccc}
%         \toprule
%         Base/Mtr & BED & PE & BE & MED & CD & BEN & STE & SHE & GE & IE \\
%         \midrule
        
% PBNS+ & - & - & - & - & - & - & - & - & - & -  \\

% HOOD+ & - & - & - & - & - & - & - & - & - & - \\

% CF+  & - & - & - & - & - & - & - & - & - & - \\

% SENC+ & - & - & - & - & - & - & - & - & - & - \\

% Ours  
% & \textbf{0.52095} & \textbf{0.05263} & 0.17482 & \textbf{14.89742} & \textbf{2.07272} & 0.03027 & \textbf{0.00261} & \textbf{0.01214} & \textbf{0.21217} & 0.00198 \\
%         \bottomrule
%     \end{tabular}
%     }
%     \label{tab:other_elastic}
% \end{table*}

\section{Methodology Details}

We represent a garment as a triangular mesh of $N$ vertices and $D$ edges, whose state $\mathcal{S}$ is denoted by the vertex positions $\mathbf{x} \in \mathbb{R}^{3 N}$ and velocities $\dot{\mathbf{x}} \in \mathbb{R}^{3 N}$. We denote the garment motion by $\{\mathcal{S}^{(t)} | t \in \mathbb{Z}, t \in [0, T]\}$, where $\mathcal{S}^{(t)} = \{\mathbf{x}^{(t)}, \dot{\mathbf{x}}^{(t)}\}$ is the garment state at time step $t$. We define the body state $\mathcal{B}$ by a joint feature vector $\mathbb{R}^9$ containing the joint location and orientation~\cite{bertiche2022neural}, and denote the skeleton state at time $t$ as $\mathcal{B}^{(t)} \in \mathbb{R}^{9 j}$ where $j$ is the number of joints. To represent the human body surface with respect to the skeleton, we use SMPL~\cite{loper2015smpl}.

Our simulator predicts the future garment states, given the garment initial state and a body motion: $\mathcal{S}^{(1:t)} = f_{sim}(\mathcal{S}^{(0)}, \mathcal{B}^{(0:t)}; \boldsymbol{\Theta})$ where $f_{sim}$ denotes our simulator which learns $\boldsymbol{\Theta}$ through minimizing our novel energy. Following existing research where physics-based simulations are equivalent to optimizations over a physical system's energy~\cite{martin2011example}, our method also turns simulation into optimization. This optimization seeks the garment states with low potential energy which aims to satisfy the physical constraints encoded in the loss function:
\begin{equation}
    \mathcal{L} = \sum_{t}^T \frac{h^2}{2} (\frac{\frac{\mathbf{x}^{(t+1)} - \mathbf{x}^{(t)}}{h} - \dot{\mathbf{x}}^{(t)}}{h})^\top \mathbf{M} (\frac{\frac{\mathbf{x}^{(t+1)} - \mathbf{x}^{(t)}}{h} - \dot{\mathbf{x}}^{(t)}}{h}) + W^{(t)}
    \label{eq:app_loss_all}
\end{equation}
where $\mathbf{M}$ is the garment lumped mass matrix, $W$ denotes the potential energy due to \eg, gravity, garment stretching energy, bending energy, \etc, and $h$ is the time step size. This minimization is equivalent to using implicit Euler~\cite{baraff2023large} to solve the equation defined by the Newton's Second law: $\mathbf{M}\ddot{\mathbf{x}}=\mathbf{f}=-\nabla W$, where $\mathbf{f}$ is the resultant force imposed on the garment~\cite{martin2011example}.

\begin{figure*}[tb]
    \centering
    \includegraphics[width=0.9\linewidth]{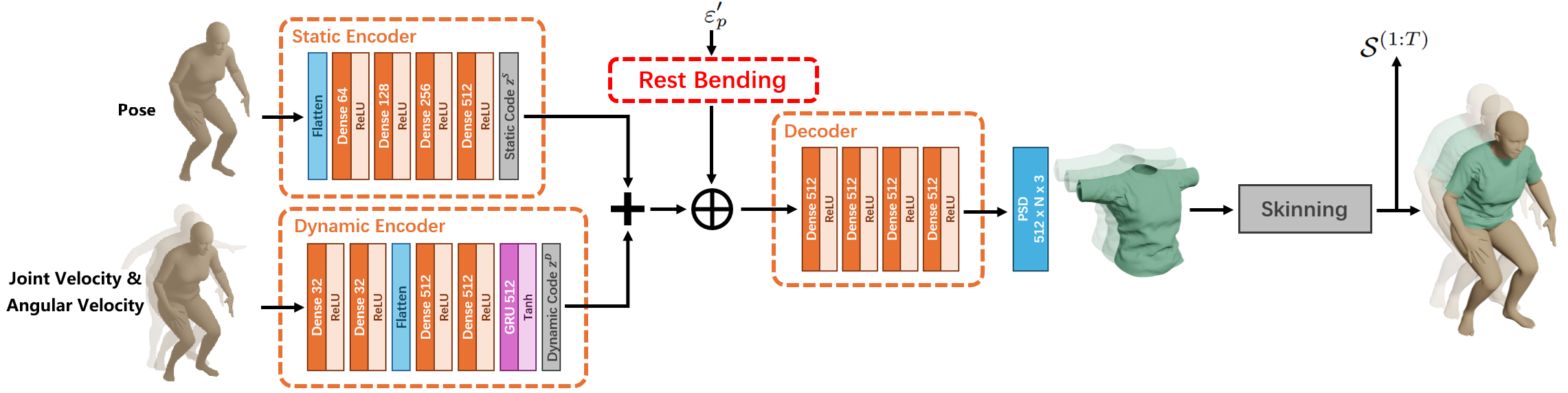}
    \caption{Elastic Net (E-Net): The input consists of a pose and its first-order time derivative. After passing through the encoder, the RB predicted by P-Net is used as a conditional input to the decoder, which then predicts the final garment deformation $\mathcal{S}^{(1:T)}$. Based on the predicted deformation, the RB is computed and subsequently used for P-Net training.
}
    \label{fig:app_elastic_net}
\end{figure*}

\begin{figure*}[tb]
    \centering
    \includegraphics[width=0.9\linewidth]{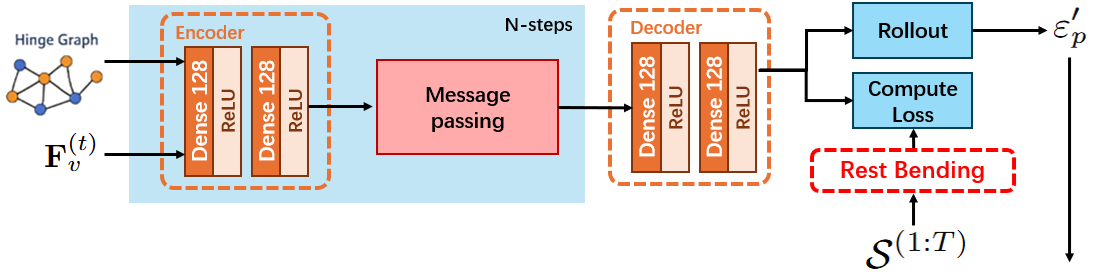}
    \caption{Plastic Net (P-Net): Hinge-level node and edge features at time step t are first embedded by MLP encoders. We then apply 3 message-passing steps, where edge features are updated using an edge MLP 
    $f_{v\to e}(e_{ij}, v_i, v_j)$ and node features are updated using a node MLP 
    $f_{e\to v}(v_i, \mathrm{mean}_j(e_{ij}))$. After decoding, the Rest Bending (RB) computed by \cref{eq:app_delta_plastic,eq:app_epsilon_plastic} from the E-Net predicted deformation $\mathcal{S}^{(1:T)}$ as the training target for learning. By rolling out P-Net over time produces the predicted  $\varepsilon'_p$ which is used as RB for condition E-Net.}
    \label{fig:app_plastic_net}
\end{figure*}

\subsection{Elastic Net}

As shown in~\cref{fig:app_elastic_net}, E-Net is composed of a static encoder and dynamic encoder, which are followed by a decoder and a PSD (Pose Space Deformation~\cite{bertiche2021deepsdautomaticdeepskinning}) projection layer. The static encoder is a stack of four fully connected layers (FCLs) with hidden dimensions 64, 128, 256, and 512, respectively. We use ReLU as the activation function between the FCLs to ensure our model's nonlinearity. The dynamic encoder contains two initial 32-dimensional FCLs (ReLU, no bias), a flattening stage, two additional 512-dimensional fully connected layers (ReLU, no bias), and a GRU (Gated Recurrent Unit) with 512 units using standard tanh activations. The Rest Bending predicted by P-Net is concatenated with the fused representation of the two encoders. The decoder is a four-layer fully connected network, all with 512 hidden units and ReLU activations. Finally, a PSD layer applies a linear projection using a learnable tensor of size $[512,n,3]$ to produce per-vertex 3D outputs.

\subsection{Plastic Net}

As shown in~\cref{fig:app_plastic_net}, P-Net operates on a hinge graph derived from the garment mesh and follows an encode–process–decode message-passing architecture.
We construct an intrinsic hinge graph $\mathcal{G}=(\mathcal{V},\mathcal{E})$ from the garment mesh. Each node $v\in\mathcal{V}$ corresponds to an interior mesh hinge, defined as a pair of adjacent faces sharing a mesh edge. An edge $(v_i,v_j)\in\mathcal{E}$ is created when two hinges are adjacent in the mesh connectivity.
Plastic bending variables are defined per hinge. Message passing is therefore performed purely on this intrinsic hinge graph and does not depend on world-space proximity.
P-Net adopts an encode–process–decode Graph Neural Network similar to GraphNet-style architectures. Node and edge features are first embedded into a latent space using multilayer perceptron (MLP) encoders consisting of two fully connected layers with 128 hidden units and ReLU activations.
We then perform $3$ steps of message passing on the hinge graph. In each step, edge features are first updated using an edge MLP
$e_{ij} \leftarrow f_{v\to e}(e_{ij}, v_i, v_j)$,
followed by node updates using a node MLP
$v_i \leftarrow f_{e\to v}(v_i, \mathrm{mean}_j(e_{ij}))$.
After message passing, a decoder MLP with the same structure (two 128-unit layers with ReLU) outputs a scalar prediction per hinge.
At time step $t$, each hinge node is associated with an 8-dimensional feature vector
$\mathbf{F}_v^{(t)}=
\Big[(\varepsilon_p^{(t-1)})_v,\,
(\varepsilon_p^{(t)})_v,\,
\mathbf{m}_v^{t},\,
\mathbf{m}_v^{0}\Big]$.
$(\varepsilon_p^{(t-1)})_v$ and $(\varepsilon_p^{(t)})_v$ denote the plastic strain at the previous and current time steps.
$\mathbf{m}_v^{t}$ denotes the hinge midpoint under the E-Net predicted deformation.
$\mathbf{m}_v^{0}$ denotes the hinge midpoint in the rest configuration.
Plastic strain is defined per hinge and associated with hinge nodes in the graph by averaging incident hinge values.
Each edge in the hinge graph is associated with a 12-dimensional feature vector constructed from: relative hinge midpoint positions (in both deformed and rest configurations), timestep information, material parameters (bending coefficient and Lamé parameters).
These edge features follow the same construction used in the deformation network (E-Net).
The decoder predicts a scalar increment representing the normalized plastic strain update for each hinge.

\subsection{Loss Function}
\label{sec:app_lossTerms}
Since our method is self-supervised, we define the loss function as a physical energy, mimicking the behavior of physics-based cloth simulators~\cite{martin2011example,baraff2023large}. Essentially, optimizing the network is equivalent to seeking garment states that satisfy the physical constraints imposed by the garment mechanics and body-garment interactions. The loss function is defined as:
\begin{equation}
\label{eq:app_totalEnergy}
    W = w_{b} W_{bend} + w_{st} W_{stretch} + w_{sh} W_{shear} +  w_{c} W_{collision} + W_{gravity}
\end{equation}
where $W$ denotes the different energy terms and $w$ denotes their corresponding weights. Since existing research shows that persistent wrinkles are mainly generated by bending~\cite{narain2013folding,gong2025cloth}, we focus on the bending energy and employ existing formulations in~\cite{bertiche2022neural}.

Our differentiable plastic rule for simulating perfect plasticity is defined as:
\begin{align}
    \Delta\varepsilon^{(t)} &= \varepsilon^{(t)}-\varepsilon^{(t-1)}_{p}-\varepsilon_y 
    \label{eq:app_delta_plastic} \\
    \varepsilon^{(t)}_p &= \varepsilon^{(t-1)}_p+\text{sigmoid}(k_p\Delta\varepsilon^{(t)})\Delta\varepsilon^{(t)} 
    \label{eq:app_epsilon_plastic}
\end{align}
where $\varepsilon^{(t)}$ is the total strain of an edge at time $t$. $\varepsilon^{(t)}_{p}$ is the plastic strain at time $t$. $\varepsilon_y$ is the yield strain. $k_p$ is a coefficient.  Our bending energy is computed as:
\begin{equation}
    W_{bend}  = \frac{1}{TD} \sum_{t}^{T} \sum_{d}^{D} k_b \frac{l^2}{8a} (\varepsilon^{(t)}_d - \varepsilon^{(t)}_{p, d})^2 
    \label{eq:app_bending_loss}
\end{equation}
where $k_b$ is the bending stiffness coefficient, $l$ is the length of the edge, $a$ is the sum of the areas of the two adjacent triangles, $d$ is the edge index. $\varepsilon^{(t)}_{p, d}$ is the RB at time $t$ for edge $d$. $D$ is total number of edges in the garment. 

Stretching measures the deviation of the deformation gradient from the rest configuration:
\begin{equation}
W_{\text{stretch}} = 
\frac{1}{T} \sum_{t}^{T} \sum_{f}^{F} k_{stretch} A_f \, \bigl( \| \mathbf{J}^{(t)}_f \| - 1 \bigr)^2,
\label{eq:app_stretch_loss}
\end{equation}
where $k_{stretch}$ is the stretch stiffness coefficient, $F$ is the number of the mesh faces, $A_f$ is the area of the $f$-th triangle, and $\mathbf{J}_f \in \mathbb{R}^{2 \times 3}$ represents the deformation gradient of face $f$, and $\|\mathbf{J}_f\|$ denotes the Frobenius norm of $\mathbf{J}_f$. The shear energy captures the angular distortion (non-orthogonality) between the two local edge vectors of a triangle:
\begin{equation}
W_{\text{shear}} =
\frac{1}{T} \sum_{t}^{T} \sum_{f}^{F} 
k_{shear} A_f ( \mathbf{J}^{(t)}_{f,0} \cdot \mathbf{J}^{(t)}_{f,1} )^2,
\label{eq:app_shear_loss}
\end{equation}
where $k_{shear}$ is shear stiffness coefficient, $\mathbf{J}_{f,0}$ and $\mathbf{J}_{f,1}$ are the first and second column vectors of $\mathbf{J}_f$. $\mathbf{J}_{f,0} \cdot \mathbf{J}_{f,1}$ is the inner product between them, indicating the deviation from orthogonality.

The collision energy penalizes penetrations by adding repulsive force to push garment vertices apart when they are close to the body mesh:
\begin{equation}
    W_{collision} = \frac{1}{T} \sum_{t}^T \sum^{N}_{i} k_c \min(f_{dis}(\mathbf{x}^{(t)}_{i}) - \epsilon, 0)^2
    \label{eq:app_collision_loss}
\end{equation}
where $f_{dis}(\mathbf{x}_t)$ measures the signed distance between a garment vertex position and the body mesh, N is the number of vertices, $\epsilon$ defines the collision threshold, and $k_c$ controls the strength of the repulsive force. 

The gravity loss defines the gravity so that the garment can drapes downward naturally:
\begin{equation}
    W_{gravity} = \frac{1}{T} \sum_{t}^{T} \sum_{i}^{N}
m_i \mathbf{g}^\top \mathbf{x}_i^{(t)}
    \label{eq:app_gravity}
\end{equation}
where $\mathbf{g}$ denotes the surface gravity of the Earth, i.e., [0.0, 0.0, 9.8] $m/s^2$, and $m_i$ is computed as
\begin{equation}
\begin{aligned}
m_i=\sum_{f\in\mathcal{N}(i)}\frac{\rho}{3}A_f,
\end{aligned}
\label{eq:app_mass_area}
\end{equation}
where $\mathcal{N}(i)$ denotes the set of triangular faces incident to vertex $i$, and $\rho$ is the material density.

\section{Training Details}

\noindent\textbf{Environment} 
We implemented our simulator in TensorFlow~\cite{tensorflow2015whitepaper} with CUDA acceleration. All experiments were run on a workstation equipped with an Intel Xeon Silver 4216 CPU and an NVIDIA TITAN RTX GPU. A full training cycle takes approximately 70 hours, and the inference time is about 0.106 seconds per frame. Training requires 9 GB of GPU memory and 26.4 GB of system RAM for E-Net, and 0.6 GB of GPU memory and 2.8 GB of system RAM for P-Net.

\noindent\textbf{Key Parameters}
To ensure consistent material properties across garments, we use the same material parameters for all garments. In \cref{eq:app_totalEnergy}, the weights of the loss terms are set as $w_{b}=4\times10^{-3}$, $w_{st}=10.0$, $w_{sh}=1.0$, and $w_{c}=10.0$. In \cref{eq:app_delta_plastic,eq:app_epsilon_plastic}, $\varepsilon_y = 0.52~\text{rad}$ and $k_p = 10.0$. In \cref{eq:app_bending_loss,eq:app_stretch_loss,eq:app_shear_loss,eq:app_collision_loss}, we use $k_b = 3.96\times10^{-5}$, $k_{stretch} = 1.0$, $k_{shear} = 1.0$, $k_{c} = 5\times10^{3}$, and $\epsilon = 1\times10^{-3}~\text{m}$. In \cref{eq:app_gravity}, $\rho = 0.2$ kg/m$^2$. The training data are sampled at 30 frames per second (FPS) with a simulation time step of $h = 1/30$ s.
The E-Net is trained for 100 epochs with a batch size of 64. For P-Net, we randomly sample 2000 garment–motion time steps from the deformations predicted by E-Net, where each sample corresponds to a randomly selected garment–motion pair. The network is trained for 50 epochs with a batch size of 1. Both networks are optimized using the Adam optimizer with a learning rate of $1\times10^{-4}$.

\noindent\textbf{Data Split}
We use the AMASS dataset, with 85\% used for training, 5\% for validation, and 10\% for testing. To compare against PBS, we further select six representative motions from the test set: waist bending, ascending stairs, dancing, dodging, jumping, and raising hands. Body motion is represented using the SMPL model. Training is performed on the default female body shape, and testing is conducted on thin and heavy body-shape variants to assess generalization. We also evaluate cross-garment generalization by training P-Net on a T-shirt and testing on trousers, vest, and long-sleeve garments.

\end{document}